\PassOptionsToPackage{dvipsnames,table}{xcolor}
\documentclass[twocolumn]{article}
\usepackage[margin=0.6in]{geometry}

\usepackage{tikz}
\usepackage{amsmath}

\usepackage{amsfonts}
\usepackage{amsmath}
\usepackage[]{hyperref}
\usepackage{tcolorbox}
\usepackage{xcolor}
\usepackage{xspace}
\usepackage[ruled]{algorithm2e}
\usepackage{algpseudocode}
\usepackage{graphicx}
\usepackage{tabularx}
\usepackage{booktabs}
\usepackage{listings}
\usepackage{multirow}
\usepackage{array}
\usepackage{adjustbox}
\usepackage{dcolumn}
\usepackage{subcaption}

\newcommand{\tool}{\work{FrameShift}\xspace}

\definecolor{tblue}{RGB}{0, 90, 181}
\definecolor{tred}{RGB}{220, 50, 32}
\definecolor{dkred}{RGB}{150, 0, 0}

\newcommand{\work}[1]{\textsc{#1}}
\lstdefinestyle{cstyle}{
language=c,
basicstyle=\ttfamily\bfseries\small,
  morekeywords={virtualinvoke},
  keywordstyle=\color{blue},
  ndkeywordstyle=\color{red},
  commentstyle=\color{dkred},
  stringstyle=\color{ForestGreen},
  numbers=left,
  breaklines=true,
  numberstyle=\ttfamily\footnotesize\color{gray},
  stepnumber=1,
  numbersep=10pt,
  backgroundcolor=\color{white},
  tabsize=4,
  showspaces=false,
  showstringspaces=false,
  xleftmargin=.23in,
  captionpos=b,
  escapeinside={$}{$}
}

\newcommand\lt[1]{{\lstinline!#1!}}
\newcommand{\customsection}[1]{%
  \vspace{0.5em}
  \noindent\textbf{#1.}\xspace
}

\newcommand{\inlinecircle}[1]{%
            \tikz[baseline=(char.base)]{
              \node[shape=circle,draw,inner sep=1pt,fill=black] (char) {\textcolor{white}{#1}};
            }%
          }

\newcommand{\whitecircle}[1]{%
            \tikz[baseline=(char.base)]{
              \node[shape=circle,draw,inner sep=1pt,fill=white] (char) {#1};
            }%
          }

\begin{document}

\date{}

\title{\tool: Learning to Resize Fuzzer Inputs Without Breaking Them}

\author{
  Harrison Green\\
  \texttt{harrisog@cmu.edu}\\
  Carnegie Mellon University
  \and
  Claire Le Goues\\
  \texttt{clegoues@cmu.edu}\\
  Carnegie Mellon University
  \and
  Fraser Brown\\
  \texttt{fraserb@cmu.edu}\\
  Carnegie Mellon University
}

\maketitle

\begin{abstract}

Coverage-guided fuzzers are powerful automated bug-finding tools.
They mutate program inputs, observe coverage, and save
any input that hits an unexplored path for future mutation.
Unfortunately, without knowledge of input formats---for
example, the relationship between formats' data fields and
sizes---fuzzers are prone to generate destructive \textit{frameshift}
mutations.
These time-wasting mutations yield malformed inputs that are rejected by the target
program.
To avoid such breaking mutations, this paper proposes a novel, lightweight
technique that preserves the structure of inputs during mutation
by detecting and using \textit{relation fields}.

Our technique, \tool, is simple, fast, and does not require additional instrumentation
beyond standard coverage feedback.
We implement our technique in two state-of-the-art fuzzers, \work{AFL++} and \work{LibAFL}, and perform a 12+ CPU-year fuzzer evaluation, finding that \tool improves the performance of the fuzzer in each configuration, sometimes increasing coverage by more than 50\%. Furthermore, through a series of case studies, we show that our technique is versatile enough to find important structural relationships in a variety of
formats, even generalizing beyond C/C++ targets to both Rust and Python.

\end{abstract}

\section{Introduction}

Fuzzing is an effective tool for exploring the state space of programs and finding bugs. While the earliest fuzzer simply fed random data into UNIX programs~\cite{miller1990empirical} (and, again, found bugs!), coverage-guided fuzzers like \work{AFL++}~\cite{afl++}, and \work{LibAFL}~\cite{libafl} use mutations and feedback to explore targets more efficiently.
These coverage-guided fuzzers maintain a growing corpus of inputs. They pick inputs from this corpus, apply random mutations (bitflips, arithmetic operations, insertions from a dictionary, etc.), and then measure feedback like edge or block coverage. The fuzzers retain mutated inputs that reach new coverage, and discard mutated inputs that don't.

These fuzzers are versatile enough to find bugs in a wide variety of targets,
and, as a result, they've been widely adopted in industry.
For example, coverage-guided fuzzers have found more than 13,000 vulnerabilities across 1,000+ open-source projects as part of Google's \work{OSS-Fuzz}.\footnote{https://github.com/google/oss-fuzz}

Unfortunately, even the best modern fuzzers struggle to successfully mutate certain types of input structures.
Many common fuzz targets operate over serialized binary formats whose \emph{metadata}---e.g., \texttt{size}
and \texttt{offset} fields---describes the layout of associated data buffers.
Security-critical applications process such structured inputs:
Hardware security chips, for example, operate on \texttt{TPM} packets,
and \texttt{openssl} (and others) use \texttt{DER}-encoded
\texttt{ASN.1} messages; both formats contain multiple
nested size fields, which make them notoriously hard to mutate~\cite{discord2024:eqv_tlv}.

In general, these sorts of fields occur in almost every serialized binary format---in codecs (\texttt{PNG}, \texttt{JPEG}, \texttt{MP3}, \texttt{OGG}, etc.), document formats (\texttt{PDF}, \texttt{XLSX}, \texttt{DOCX}, etc.), cryptographic protocols (\texttt{TLS}, \texttt{SSH}), object formats (\texttt{ELF}, \texttt{PE}, \texttt{Mach-O}), and many more.
Fundamentally, any multi-part, variable-length binary data format
\textit{requires} metadata to describe its structure and delineate field boundaries.

These metadata-rich formats pose a challenge to modern fuzzers.
When a fuzzer mutates specific parts of an input---like a variable sized data buffer---without correspondingly updating related parts of the input---like the size or offset fields describing that buffer---it renders the input structurally invalid.
We call such destructive mutations \emph{frameshifts}.
As a result of the frameshift, the target program will mishandle the input data
or abort early with a validation error.
Frameshifts cause fuzzers to get stuck exploring the space of invalid inputs
and the space of inputs with the same sized structures as appear in the seed corpus;
they are unable to  discover new, interesting inputs that contain resized or shifted data. 

Existing approaches to this problem either (1)~augment the fuzzer with an input specification, allowing it to understand and generate the expected structure~\cite{aflsmart,formatfuzzer}, or (2)~learn important structures automatically during fuzzing using e.g., static analysis~\cite{tiff,nestfuzz}, coverage-guided feedback~\cite{weizz,profuzzer}, or (recently) a combination of static analysis and machine learning~\cite{aifore}.

Unfortunately, existing techniques in the first
category require manual effort and, simultaneously, risk
over-constraining the fuzzer.
Techniques in the second category don't solve the general
frameshift problem either. 
Both \work{AIFORE}~\cite{aifore} and \work{ProFuzzer}~\cite{profuzzer} are closed source, which makes widespread adoption impossible.
Further, all of \work{TIFF}~\cite{tiff}, \work{WEIZZ}~\cite{weizz}, \work{ProFuzzer}~\cite{profuzzer}, and \work{AIFORE}~\cite{aifore}
try to discern whether bytes are e.g. an integer, or an enum, or a string, etc.
They may identify relation fields, but do not attempt to identify the relationship between these fields and their target buffers---and thus cannot perform validity-preserving resizing mutations.

Further, modern fuzzers include a (rough) generalization
of the techniques prior work uses to e.g., discern valid
\lt{enum} options, byte ranges, etc.
\work{AFL++} and \work{LibAFL}, for example, use sophisticated
strategies like compare-logging (an adaptation of \work{RedQueen}~\cite{redqueen}),
which attempts to find special values and inject them into the input. 
This type of technique can, for example, identify alternative \lt{enum} options (by instrumenting switch cases), or required magic bytes (by instrumenting e.g. \lt{memcmp}), thus subsuming less general analyses that do (some of) the same thing.
Unfortunately, these features are not always enabled by default, rendering them absent from some academic evaluations~\cite{nestfuzz,aifore}---and potentially underselling the actual performance of modern general-purpose fuzzers.
In our work, we use industry-standard configurations (following \work{FuzzBench}~\cite{fuzzbench}) and focus on designing a system that confers an actual, significant benefit over state-of-the-art baseline fuzzers, even in their optimal configurations.

This paper presents a new approach to fuzzing structured
input formats by discovering relation fields and using them for structure-aware resizing mutations that preserve input validity. 
Our approach, \tool, is built on two key insights. 
First, coverage loss between a seed and a mutated input indicates that the fuzzer \emph{may} have mutated an important
relation field (e.g., size field) without mutating the corresponding data.
This indicates a \emph{potential} destructive frameshift
that can be identified dynamically, over the course of a coverage campaign.
There are, however, other reasons---reasons beyond frameshifts---that a mutation can lead to coverage loss.
For example, mutations to enums may redirect execution to a different code path, or mutations to checksums may cause the target to abort early.
Thus, our second insight is that, to identify \emph{true}
frameshifts, \tool can conduct experiments to find points where resizing a buffer restores coverage with
respect to the original destructive mutation. 
We prune the search space of potential corrective mutations using
domain-specific heuristics that let the analysis run in mere seconds
per new input.
Finally, \tool uses its newly-discovered relations to inform
fuzzing with existing mutators---thus doing structure-aware fuzzing
that avoids destructive frameshifts. 

We designed \tool to be fast, easy to integrate, and compatible with
modern fuzzers.
It can be applied to a wide range of existing fuzzers because,
inspired by prior work~\cite{weizz,profuzzer}, 
it doesn't require instrumentation beyond coverage feedback.
It also does not require manual format specifications.
Finally, \tool is performant: on unfriendly targets, the analysis
almost never tanks fuzzer performance by incurring serious overhead;
on friendly targets, it helps a fuzzer achieve new coverage quickly.

\customsection{Contributions}
We show that \tool is:
\begin{itemize}
\item \textbf{Effective} compared to industry-leading, state-of-the-art fuzzers.
  It increases coverage by an average of 6\%---and more than 50\% in certain configurations---while only suffering a 5.5\% coverage loss (on average) for the worst-case target.
\item \textbf{Versatile}. Unlike static analysis-based approaches, it is not limited to C/C++:
  we run \tool out-of-the-box on Rust and Python fuzz harnesses.
\item \textbf{Capable} of identifying real, target-specific size and offset fields---even nested ones---in a variety of popular binary formats.
It even discovers semantic differences in input formats across two programs that parse the same input type. 
\end{itemize}

\section{Overview}
\label{sec:overview}

This section describes the challenges that fuzzers face with structured
inputs (Section~\ref{sec:motivation}),
and then outlines the intuition behind our approach
(Section~\ref{sec:intuition}).
We use the \texttt{TPM} format as a running example by
investigating the \texttt{ms-tpm-20-ref} target,\footnote{https://github.com/microsoft/ms-tpm-20-ref}
a reference TPM 2.0 specification and simulator developed by Microsoft.
 
\subsection{Motivating Example}
\label{sec:motivation}

\begin{figure}
  \centering
  \resizebox{\linewidth}{!}{
    \includegraphics{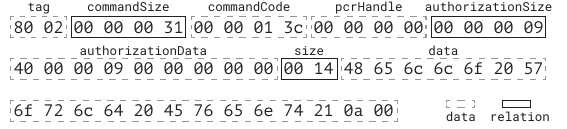}
  }
  \caption{An example \lt{TPM_PCR_Event} packet with annotated fields.}
  \label{fig:tpm_packet}
\end{figure}

\begin{figure}
  \centering
  \resizebox{\linewidth}{!}{
    \includegraphics{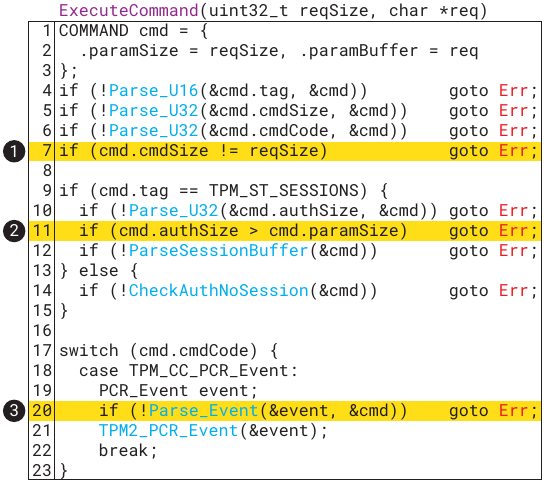}
  }
  \caption{Example parsing logic for a \lt{TPM_PCR_Event} command in \texttt{ms-tpm-20-ref} (rewritten for clarity). Three size validation checks are highlighted.}
  \label{fig:tpm_code}
\end{figure}

\autoref{fig:tpm_packet} shows a \lt{TPM_PCR_Event} command
packet (command code \lt{0x13c}) with the payload ``Hello World Event!''.
The parsing and command execution logic
(\autoref{fig:tpm_code},
simplified for presentation) contains a number of
interesting edge cases for a fuzzer to discover.  It parses fields
from the data buffer (lines 4-7); validates authorization data, or follows an
alternative path (\lt{CheckAuthNoSession}) if there exists no authorization
session (lines 9-15); and then invokes the correct execution handler (lines
17-22).  If any of the parser checks fail, execution terminates with error
handling (\lt{goto Err}).  

The best outcome for a fuzzer for this code
snippet involves reaching
the call to \lt{TPM2_PCR_Event} (line 21) with many different inputs.  
To do so, the fuzzer must navigate three nested size fields that must stay
synchronized with one another and with the described data:
\lt{cmdSize} must match the total packet size (check
\inlinecircle{1});
\lt{authSize} must fit within the remaining data (check \inlinecircle{2});
and payload event \lt{size} (drawn from the payload data, interpreted as a
variable size \lt{PCR_Event} buffer) must equal the size of the remaining
\lt{data} field (check \inlinecircle{3}). 

It is therefore \emph{extremely difficult} for a fuzzer to mutate an existing
valid TPM message into another valid message with a different payload
size.  Any mutation that changes the message size must also change
\lt{cmdSize}; insertions or deletions in
\lt{authData} must also update \lt{authsize}; and attempts to resize the
\lt{data} buffer must also update \lt{size}.

To demonstrate the challenge this poses, we ran five state-of-the-art fuzzers
(along with our own 2 \tool variants) on the \texttt{ms-tpm-20-ref} target using
the example TPM packet from \autoref{fig:tpm_packet} as the seed input.
Each fuzzer ran for 48 hours for 10 repetitions (see \autoref{sec:setup} for the full
experimental setup). We then analyzed the resulting corpus to see how frequently
each fuzzer was able to find \textit{newly sized} (i.e. differing in
\lt{cmdSize}, \lt{authSize}, or \lt{size} from the seed) inputs that
passed each of the highlighted validation checks.

\autoref{table:tpm_case_study} shows results.  \emph{None} of the
state-of-the-art fuzzers were able to find a \emph{single} newly sized input
that got to---and passed!---check \inlinecircle{3}. All generated inputs that \emph{did} pass that check had an
\lt{authSize} of 9 bytes and a data payload \lt{size} of exactly
20 bytes, like the seed.
These fuzzers were effectively stuck,
unable to successfully perform a resizing mutation even after 20 CPU-days of
fuzzing.\footnote{\work{WEIZZ} and \work{NestFuzz} do identify some structures, but are not able to identify and preserve these size fields.}

Our \tool variants, under the same configuration, were able to find 14 and 8 newly sized \lt{TPM_PCR_Event} commands respectively, unlocking new codepaths in the \lt{TPM2_PCR_Event} handler. Furthermore, they were both able to find an order of magnitude more newly sized inputs reaching the prior checks, discovering many more command types in the process.
As a result, \tool variants found an average of 15.3\% more coverage than the baseline fuzzers on \texttt{ms-tpm-20-ref} in a 48-hour fuzzing campaign (\S\ref{sec:rq2}).

\begin{table}
  \centering
  \begin{tabularx}{\linewidth}{p{0.1em}Xrrrr}
    \toprule
    & & & \multicolumn{3}{c}{\textbf{New Variants}} \\
    \cmidrule(l){4-6}
    \multicolumn{2}{l}{\textbf{Fuzzer}} & \textbf{Corpus Size} & \inlinecircle{1} & \inlinecircle{2} & \inlinecircle{3} \\
    \midrule
    \multicolumn{3}{l}{\textit{State-of-the-art Fuzzers}} \\
    & \work{AFL++}~\cite{afl++} & 17062 & 41 & 36 & 0 \\
    & \work{LibAFL}~\cite{libafl} & 23360 & 52 & 53 & 0 \\
    & \work{AFL}~\cite{afl} & 15694 & 35 & 38 & 0 \\
    & \work{WEIZZ}~\cite{weizz} & 24231 & 16 & 15 & 0 \\
    & \work{NestFuzz}~\cite{nestfuzz} & 11593 & 28 & 32 & 0 \\
    \midrule
    \multicolumn{3}{l}{\textit{With \tool (our work)}} \\
    & \work{AFL++(FS)} & 21693 & 359 & 399 & 14 \\
    & \work{LibAFL(FS)} & 37255 & 194 & 287 & 8 \\
    \bottomrule \\
  \end{tabularx}
  \caption{Newly sized variants passing the highlighted validation checks in \texttt{ms-tpm-20-ref} with and without \tool after 48 hours of fuzzing and 10 repetitions.}
  \label{table:tpm_case_study}
\end{table}

The \texttt{TPM} example illustrates a larger issue that affects
a wide range of binary formats (\texttt{ELF}, \texttt{PNG}, \texttt{ASN.1}, etc.)
that are essential in the modern software stack. 
Our approach, which we describe next, overcomes this issue
by allowing fuzzers to automatically find relation fields in input formats and use them to enable structure-aware resizing mutations.

\subsection{\tool Intuition}
\label{sec:intuition}
\begin{figure*}
  \centering
  \resizebox{\linewidth}{!}{
    \includegraphics{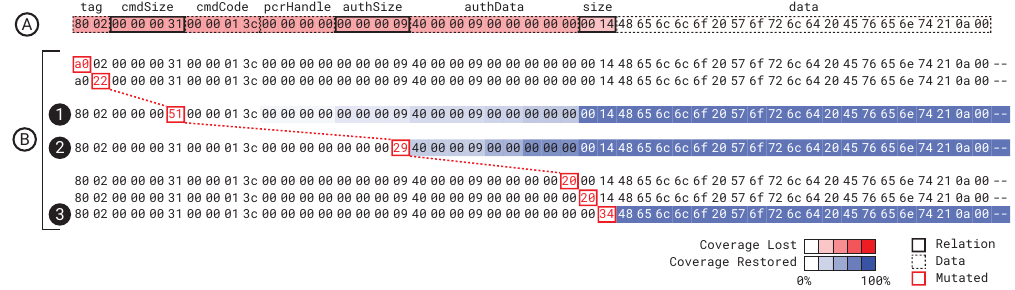}
  }
  \caption[Illustration of the double-mutant experiment on a \lt{TPM_PCR_Event} packet. (A): The original input; bytes that cause a loss-of-coverage when mutated are highlighted red. (B): Each row shows a mutated byte that caused a loss-of-coverage (boxed in red) along with every insertion point that was able to restore coverage (highlighted in blue). Some rows with no coverage restoration are omitted for brevity (indicated by the dotted red line). Rows (1), (2), and (3) correspond to the discovery of the highlighted validation checks in \autoref{fig:tpm_code}.]{Illustration of the double-mutant experiment on a \lt{TPM_PCR_Event} packet. \whitecircle{A}: The original input; bytes that cause a loss-of-coverage when mutated are highlighted red. \whitecircle{B}: Each row shows a mutated byte that caused a loss-of-coverage (boxed in red) along with every insertion point that was able to restore coverage (highlighted in blue). Some rows with no coverage restoration are omitted for brevity (indicated by the dotted red line). Rows \inlinecircle{1}, \inlinecircle{2}, and \inlinecircle{3} correspond to the discovery of the highlighted validation checks in \autoref{fig:tpm_code}.}
  \label{fig:tpm_display}
\end{figure*}

Our technique identifies structural metadata---the
position and target of each size field in the TPM input packet---and augments a fuzzer to preserve
the relationships between that metadata automatically during mutation.
The key idea is to (1) identify \emph{potential}
relation fields by observing mutations that cause coverage loss, and to (2)
validate \emph{true} relation fields by experimenting with new
mutations that restore coverage by changing data size. 
Such ``double-mutants'' that preserve coverage
while resizing the input indicate a likely relation field.

\autoref{fig:tpm_display} illustrates a walkthrough of \tool's \emph{double-mutation}
experiments on our example \lt{TPM_PCR_Event} packet; we discuss
each step of the process next.

\customsection{\whitecircle{A} Disrupting Coverage}
The first key observation\footnote{Which \work{ProFuzzer}~\cite{profuzzer} also relies on.} is that it's possible
to \emph{indirectly} identify validation checks
(without e.g., statically searching for them ahead of time).
At runtime, our tool can use loss-of-coverage
as evidence of a validation check.
For example, starting with a valid seed and mutating the \lt{cmdSize} field will
result in a new input that fails to reach some of the originally-covered program
bits (those after check \inlinecircle{1} in \autoref{fig:tpm_code}).

In \autoref{fig:tpm_display}, row \whitecircle{A} highlights the bytes for
which incrementing the byte's value by \lt{0x20} causes a loss in
coverage (all bytes up to and including \lt{size});
none of the bytes in the \lt{data} field are highlighted, because mutating them
does not change the execution path. 
The highlighted bytes are \emph{candidate} relation fields: it is possible
that they correspond to metadata in the input format that must be synchronized
with data in the input.

\customsection{\whitecircle{B} Restoring Coverage}
Given these candidate relations, the loss-of-coverage can be explained by
either: (1) a \textit{frameshift} mutation (what we're looking for), or (2) some
other validated part of the testcase (uninteresting for our purposes). For example,
mutating \lt{cmdCode} may cause the program to trigger a different command
handler.

To disambiguate these scenarios, \tool aims to automatically identify
points at which bytes can be inserted into the input to restore (most of) the
original coverage.  That is, if the fuzzer mutated a size field by incrementing its
value by $N$, there should be a point in the input where inserting $N$ bytes
``re-syncs'' the size field with the data it describes. 
Critically, this is likely only possible if the
original mutated field \emph{actually described a size.} It is unlikely, for
example, that a fuzzer can increment the \lt{cmdCode} field by $N$, and then
insert $N$ bytes elsewhere that restore the original execution behavior.

Although there are 29 bytes that can be mutated to disrupt coverage for the
\lt{TPM_PCR_Event} packet, there are only three 
for which we can find an associated \textit{insertion point} that restores some
of the original coverage (\autoref{fig:tpm_display}, rows \inlinecircle{1},
\inlinecircle{2}, and \inlinecircle{3}).\footnote{Initially, the only
identifiable byte is \lt{cmdSize}; the others are discoverable only after it
is identified.} 

Row \whitecircle{B}\inlinecircle{1}, shows that inserting right before the
\lt{pcrHandle}  restores a small percentage of coverage (lightly shaded),
but that the most coverage is obtained by inserting near the end of the file.  
Practically, this is because inserting near the end of the file preserves the
authentication section, passing checks \inlinecircle{2} and \inlinecircle{3}.
Row \whitecircle{B}\inlinecircle{2} corresponds to insertions that correct the \lt{authSize} field.
Some coverage is restored when inserting inside the
\lt{authData} region. However, this is likely to corrupt the
authentication data, passing check \inlinecircle{2} and failing
\inlinecircle{3}. Inserting anywhere after the existing authentication data
restores more coverage. Finally, part
\whitecircle{B}\inlinecircle{3} corresponds to the \lt{size} field of the
\lt{PCR_Event}, where any insertion inside the \lt{data} region restores coverage equally.

This example illustrates the intuition behind \tool's approach to dynamically
identifying relation fields. Testing every byte or insertion point
(as in this illustration) is prohibitive in practice; moreover, discovering
certain relation fields is often impossible without discovering others
(e.g., \lt{cmdSize}).  Our implementation uses heuristics to prune
the search space and makes this entire process practical (i.e., on the order of seconds or milliseconds for a single input).

\section{\tool}

In this section, we expand on the intuition from the previous section
and describe our design in detail.
We start by formalizing the concept of a \emph{relation field}, an abstraction over different types of size/offset fields (\S\ref{sec:structured-inputs}).
We then describe how to discover these relation fields, using heuristics to prune the search space of possible relations
(\S\ref{sec:relation-discovery}).
Finally, we discuss how \tool uses relations to implement
structure-aware mutations (\S~\ref{sec:structure-aware-mutation}).

\subsection{Structured Inputs}
\label{sec:structured-inputs}

Disruptive frameshift mutations occur when a fuzzer modifies input bytes that correspond
to metadata (e.g. a size field), without fixing up the corresponding data.

To avoid these frameshifts, our goal is to discover the locations of these size fields and their associated data--the part of the input they describe the size of.
For the purposes of mutation, \texttt{size} and \texttt{offset} fields can be treated the same. A \texttt{size} field represents the length of some span of the input data. An \texttt{offset} field represents the position of some part of data in relation to the start of the input, or equivalently: a \texttt{size} field that describes the length of the input before the data.

We generalize both these forms as a \textit{relation field}
$$R := (a, b, p, s, e)$$
consisting of a field at position $p$ with size $s$ and endianness $e$, that represents the length of some span of the input data. The span is defined by a start position $a$ and an end position $b$, where $a < b$.

\begin{figure}
  \centering
  \resizebox{\linewidth}{!}{
    \includegraphics{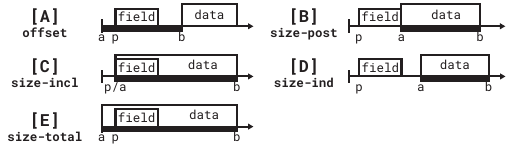}
  }
  \caption{Examples of size/offset fields encoded as relation fields.}
  \label{fig:relation_fields}
\end{figure}

This construction generalizes many types of size/offset fields, as depicted in
\autoref{fig:relation_fields}. For an offset field (A), the start position $a$ is 0 and the end position $b$ designates where the data starts (i.e., the offset). For a size field, the actual data field starts at $a$ and ends at $b$. However, the size field itself may be positioned in different ways with respect to the data it describes, i.e. immediately before it (B), just inside it (C), some other arbitrary location (D), or the size field may be the entire input (E).

\subsection{Automatic Relation Discovery}
\label{sec:relation-discovery}

The objective of the relation analysis is to discover important relation fields
in particular inputs. This analysis runs once on every new corpus input
the fuzzer discovers.
When the fuzzer attempts to mutate an input
associated with relations, is uses relation data to ``fix up''
fields whose relationships were affected by the mutation.

The analysis first identifies destructive mutations
to the input---mutations that lead to coverage loss. 
Destructive mutations indicate that the mutated
bytes \emph{might} correspond to relation fields.
The analysis then attempts \emph{restorative} mutations to undo
the destruction caused by the original mutation.
If such restoration is possible, it indicates that the first
mutation was indeed a frameshift, and that the mutated bytes correspond to
a relation field.
It also provides evidence as to the nature of the relation (i.e., what part of the input it describes).

\customsection{Destructive and restorative mutations}
A given input $I$'s coverage profile $F$ is represented abstractly as a set of bits:
$$
F(I) := \{f_1, f_2, \ldots, f_n\}
$$
Depending on the instrumentation used, these bits may represent blocks hit,
edges taken, etc.

A destructive mutation loses some of this original coverage. We set a threshold
$T_\text{loss}$ to indicate the percentage of coverage loss necessary to signal
a destructive mutation, i.e., $I^-$ is a
destructive mutation iff.:
$$
|F(I) - F(I^-)| \geq T_\text{loss} \cdot |F(I)|
$$
Conversely, a \textit{restorative} mutation restores some of the lost coverage,
and is defined by the threshold $T_\text{restore}$.
That is, mutating an input $I^-$ produces a \textit{restoring mutant} $I^+$ iff.:
$$
|F(I^+) \cap (F(I) - F(I^-))| \geq T_\text{restore} \cdot |F(I) - F(I^-)|
$$

In our experiments, $T_\text{loss} = 0.05$ and $T_\text{restore} = 0.2$. These values work well,
but per-target tuning is an interesting future direction.

\customsection{Candidate Relation Field Identification}
The first step in the analysis is to identify potential relation fields. To this
end, the analysis iterates over every field size $s \in \{8,4,2,1\}$, endianness
$e \in \{\text{big}, \text{little}\}$
and potential field position $p \in \{0, \ldots, \text{size}(i)-s\}.$
For each configuration, the analysis deserializes the input
bytes at position $p$ to obtain a value $v$. To prune the search space, it only
considers fields with a value $v \leq \text{size}(i)$ for some input $i$, since
sizes or offsets must have values that are at most the size of the input.
Given a
candidate field, the analysis mutates its value to test if
doing so causes 
coverage loss that exceeds $T_\text{loss}$. For $s > 1$, the mutation increments by \lt{0xff}, forcing a carry from the least-significant byte (which distinguishes between little- and big-endian fields); for $s = 1$, the mutation increments the value by \lt{min(0x20, 0xff-v)} such that the increase is large enough to cause potential frameshifts but does not overflow the field.
If the mutations lead to coverage loss above $T_\text{loss}$,
the mutated bytes are a candidate relation, and analysis continues with insertion point discovery.

\customsection{Insertion Point Discovery}
Just because a mutation reduces coverage doesn't mean
that the mutated bytes store a relation field.
For example, mutating important constants, checksums, or changing enum values is likely
to invalidate the input or change execution flow, degrading coverage.
Thus, the
second phase of the analysis seeks evidence that a candidate relation field corresponds
to a \emph{true} relation field; if so, it also collects evidence about the nature
of the relation.

For each candidate field, the analysis iterates over potential
insertion points---places where we can insert bytes---in search of restorative mutations.
Iterating over every byte in
the input is prohibitively expensive, as it requires invoking the target program
for every byte. Instead, we limit the search based on a smaller set of
\emph{anchor points}. Specifically, in practice, most
relation fields occur in one of the forms depicted in
\autoref{fig:relation_fields}. Thus, the \texttt{start} of the target span is
often one of $0$ (for offset / size-total fields), $p$ (for size-inclusive
fields), or $p+s$ (for size-post fields). In each of these cases, we test the
corresponding \texttt{end} position (\texttt{start} + $v$) as a candidate
insertion point. We select the insertion point that restores the most
coverage, as long as it exceeds the $T_\text{restore}$ threshold.

A slightly harder case is size-indirect form (D), where the target start
position may occur at an arbitrary point in the input. In practice, since the target program needs to be able to actually locate this data, there is often some other metadata (another size or offset field) that indicates where the start of the data should be.

For example, in \texttt{ELF} files,
there are 8-byte \texttt{size} and \texttt{offset} fields describing the
location of program and section headers. Therefore, to find these fields
efficiently, we expand the insertion point search to consider start positions at $R.p$,
$R.a$, and $R.b$ for every true relation field $R$ that we have \emph{already} discovered.
During analysis, \tool first identifies the \texttt{offset} field, and then
uses it as an anchor point to identify the \texttt{size} field insertion
point.

\tool is a fundamentally heuristic analysis:
candidate relation fields with viable insertion points are \emph{likely} (but not guaranteed) to be real relation fields.
It is possible---but rare!---that inserting bytes can restore coverage for other reasons (e.g., if the initial mutation actually modified an \lt{enum} value, but the inserted bytes coincidentally reintroduced another input structure which hit the original codepath).

\subsection{Structure-aware Mutation}
\label{sec:structure-aware-mutation}

\tool uses relation fields to augment standard byte-level mutations,
yielding \emph{structure-aware} mutations.
Next, we describe standard fuzzer mutations, and how
\tool adapts these mutations to account for relations. 

\customsection{Raw Mutations}
Standard fuzzer mutations
manipulate a 
raw input $I$ through three interfaces:
$\texttt{Replace}(I,i,V)$ (replace the subsequence starting at $i$ with $V$),
$\texttt{Insert}(I,i,V)$ (insert subsequence $V$ at position $i$), and
$\texttt{Remove}(I,i,n)$ (remove $n$ bytes after position $i$)
(\autoref{fig:unstructured-mutators}). Fuzzer mutators typically perform several of
these actions at once. For example, a \textit{splicing} mutator may perform
some combination of \texttt{Replace} and \texttt{Insert}.

\begin{figure*}[t]
  \centering
  \scriptsize
  \begin{subfigure}[t]{.49\textwidth}
    \centering
    \(
      \begin{aligned}
        \texttt{Replace}(I,i,V) &\rightarrow
          (I_0,I_1,\dots,I_{i-1})\Diamond V\Diamond
          (I_{|V|},I_{|V|+1},\dots,I_{|I|-1}) \\[2pt]
        \texttt{Insert}(I,i,V) &\rightarrow
          (I_0,I_1,\dots I_{i-1})\Diamond V\Diamond
          (I_i,I_{i+1},\dots,I_{|I|-1}) \\[2pt]
        \texttt{Remove}(I,i,n) &\rightarrow
          (I_0,I_1,\dots,I_{i-1})\Diamond
          (I_{i+n},I_{i+n+1},\dots,I_{|I|-1})
      \end{aligned}
    \)
    \caption{Unstructured mutation operators}
    \label{fig:unstructured-mutators}
  \end{subfigure}
  \hfill
  \begin{subfigure}[t]{.49\textwidth}
    \centering
    \(
      \begin{aligned}
        \texttt{Replace}(S,i,V) &\rightarrow
          (\texttt{Replace}(S.I,i,V),\;S.\mathbb{R}) \\[2pt]
        \texttt{Insert}(S,i,V) &\rightarrow
          (\texttt{Insert}(S.I,i,V),\;
            \{\texttt{OnInsert}(R,i,V)\mid R\!\in\!S.\mathbb{R}\}) \\[2pt]
        \texttt{Remove}(S,i,n) &\rightarrow
          (\texttt{Remove}(S.I,i,n),\;
            \{\texttt{OnRemove}(R,i,n)\mid R\!\in\!S.\mathbb{R}\})
      \end{aligned}
    \)
    \caption{Structured mutation operators}
    \label{fig:structured-mutators}
  \end{subfigure}

  \caption{Side-by-side comparison of unstructured (left) and structured (right) mutation operators.  $\Diamond$ denotes sequence concatenation.}
  \label{fig:mutation-operators-side}
\end{figure*}

\customsection{Structured Mutations}
We redefine these mutation operators to act on a structured input $S := (I,\mathbb{R})$ with input $I$ and set of learned relations $\mathbb{R}$
(\autoref{fig:structured-mutators}), by first applying the mutation to the
underlying input $I$ and then invoking \texttt{OnInsert}
(Algorithm~\ref{alg:oninsert}) or \texttt{OnRemove}
(Algorithm~\ref{alg:onremove}) to track which relation fields need to shift or update their value as a result of the operation.
It is important to perform this bookkeeping as fuzzers may stack these splicing mutations multiple times in a row. 
After all mutations, and before executing the test case, \tool re-serializes relation fields to apply their new values to the underlying input.

\customsection{Accommodating Havoc}
While \tool is designed to identify (and preserve the validity of) resizing mutations, we don't want to inadvertently restrict the fuzzer from making destructive mutations that would unlock new coverage. Therefore, during a mutation, if the fuzzer tries to perform an action that is incompatible with the current set of relations---e.g., inserting into the middle of a relation field itself---\tool temporarily
deletes that relation and avoids re-serializing it. Thus, \tool updates relation fields without also over-constraining the
fuzzer.\footnote{This is also important in the rare case that \tool incorrectly identifies a relation field.}

\begin{algorithm}
  \caption{OnInsert}
  \KwData{Relation $R$, index $i$, sequence $V$}
  \lIf{$i \le R.p$}{$R.p \gets R.p + |V|$}
  \lIf{$i < R.a$}{$R.a \gets R.a + |V|$}
  \lIf{$i \le R.b$}{$R.b \gets R.b + |V|$}
  \Return{R}
  \label{alg:oninsert}
\end{algorithm}

\begin{algorithm}
  \caption{OnRemove}
  \KwData{Relation $R$, index $i$, size $n$}
  \lIf{$i \le R.p$}{$R.p \gets R.p - \texttt{min}(R.p-i, n)$}
  \lIf{$i \le R.a$}{$R.a \gets R.a - \texttt{min}(R.a-i, n)$}
  \lIf{$i \le R.b$}{$R.b \gets R.b - \texttt{min}(R.b-i, n)$}
  \Return{R}
  \label{alg:onremove}
\end{algorithm}

\subsection{Implementation}
\label{sec:prototypes}

We implement the \tool algorithm in both \work{AFL++} and \work{LibAFL}, two
industry-leading fuzzers. 
Our implementations integrate with existing fuzzer mutators and require no changes to instrumentation.
These variants are
denoted as \work{AFL++(FS)} and \work{LibAFL(FS)} throughout subsequent sections.
Both implementations are open-source under a permissive license, available at \url{https://github.com/hgarrereyn/AFLplusplus-FrameShift} and \url{https://github.com/hgarrereyn/LibAFL-FrameShift}.

\customsection{\tool in \work{AFL++}}
Our \work{AFL++} implementation of \tool consists of 600 lines of C that implement a new fuzzer stage to run analysis and store relation metadata in queue inputs. \work{AFL++(FS)} tracks insertions and deletions from the havoc and splice mutators, and then re-serializes relation data before executing test cases.

\customsection{\tool in \work{LibAFL}}
For \work{LibAFL} we write a modular fuzzer stage and custom input type in roughly 1600 lines of Rust. The implementation is functionally identical to our \work{AFL++} fork but implemented in a canonical \work{LibAFL} style. As such, it is plug-and-play with many other \work{LibAFL} modules.
This lets us use \work{LibAFL}'s support for other languages, and apply \tool \emph{out of the box} to both Rust and Python targets (see Section~\ref{sec:rq4}). 

\section{Evaluation}
\label{sec:eval}
This section answers the following research questions:
\begin{itemize}
    \item \textbf{RQ1 (Performance)}: How does \tool compare to SOTA binary and
    structure-aware fuzzers at finding coverage? (Section~\ref{sec:rq1})
    \item \textbf{RQ2 (Applicability)}: Where is \tool most/least effective?
    What are the failure cases? (Section~\ref{sec:rq2})
    \item \textbf{RQ3 (Case Study)}: What types of structures can \tool identify
    in real-world targets? (Section~\ref{sec:rq3})
    \item \textbf{RQ4 (Versatility)}: How versatile is \tool with respect to
    different languages and different forms of coverage feedback? (Section~\ref{sec:rq4})
\end{itemize}

Section~\ref{sec:setup} describes benchmarks and baselines; subsequent sections
address each research question in turn.

\subsection{Experimental Setup}
\label{sec:setup}

\customsection{Benchmarks}
For the large-scale experiment (RQ1), we select 16 benchmarks
(\autoref{table:targets}): all the binary-format targets in
\work{FuzzBench}~\cite{fuzzbench}, two text-based formats (\texttt{jsoncpp} and
\texttt{libxml2}) from \work{FuzzBench}, and two more binary
formats: \texttt{ms-tpm-20-ref} and \texttt{qpdf} from
\work{OSS-Fuzz}~\cite{ossfuzz}, inspired by discussions about hard-to-fuzz file formats~\cite{discord2024:eqv_tlv}.
We include the text-based formats as a baseline for measuring
\tool's worst-case overhead (since we don't expect our technique
to work well on these benchmarks).

\begin{table}
  \centering
  \begin{tabular}{lll}
    \toprule
    Benchmark & Format & Commit \\
    \midrule
    bloaty & ELF/Mach-O/WebAssembly & \texttt{52948c1} \\
    freetype2 & TTF/OTF/WOFF & \texttt{cd02d35} \\
    harfbuzz & TTF/OTF/TTC & \texttt{cb47dca} \\
    lcms & ICC-profile & \texttt{f0d9632} \\
    libjpeg-turbo & JPEG & \texttt{3b19db4} \\
    libpcap & PCAP & \texttt{17ff63e} \\
    libpng & PNG & \texttt{cd0ea2a} \\
    ms-tpm-20-ref & TPM & \texttt{6b72d66} \\
    openh264 & H.264 & \texttt{045aeac} \\
    openssl & DER & \texttt{b0593c0} \\
    openthread & IPV6-packet & \texttt{2550699} \\
    qpdf & PDF & \texttt{2cb2412} \\
    vorbis & OGG & \texttt{84c0236} \\
    woff2 & WOFF & \texttt{8109a2c} \\
    \midrule
    jsoncpp & JSON (text) & \texttt{8190e06} \\
    libxml2 & XML (text) & \texttt{c7260a4} \\
    \bottomrule \\
  \end{tabular}
  \caption{Benchmark programs}
  \label{table:targets}
\end{table}

\customsection{Baseline Fuzzers}
We select five additional baseline fuzzers for our evaluation
(\autoref{table:fuzzers}). \work{AFL++}~\cite{afl++} and
\work{LibAFL}~\cite{libafl} act as direct baselines for our prototype
implementations. We also include \work{AFL}~\cite{afl}, since
it is the direct baseline for \work{NestFuzz}.

We also evaluate against two fuzzers that do structural inference. \work{WEIZZ}~\cite{weizz} uses coverage feedback and extra comparison instrumentation to identify structures in chunk-based binary formats. \work{NestFuzz}~\cite{nestfuzz}  models the input processing logic of a program via dynamic taint analysis to discover dependencies which are used during mutations.

We do not include \work{AIFORE}~\cite{aifore} or \work{ProFuzzer}~\cite{profuzzer} because both are closed-source.\footnote{Authors of \work{AIFORE} did not respond to our request for source code. Authors of \work{ProFuzzer} provided source code, but did not respond to our request for clarification when we could not run the tool.}
We also omit \work{TIFF}~\cite{tiff} because it requires paid decompiler software and relies on now-outdated versions of Intel Pin;
recent results suggest it would be outperformed by both \work{NestFuzz} and
\work{WEIZZ}, which we include.
These fuzzer baselines represent the state-of-the-art both in general purpose coverage-guided fuzzing and automated binary structure-aware fuzzing.

\customsection{Fuzzer Configurations}
For our prototype and all baseline fuzzers except \work{NestFuzz}, we use the \work{FuzzBench} configuration.
In both \work{LibAFL} and \work{AFL++}, this configuration includes
\work{REDQUEEN}-style compare-logging~\cite{redqueen} and
dictionaries, two features that work well in practice~\cite{liu2023sbft}.
Our tool variants are configured identically to \work{AFL++} and
\work{LibAFL}, except they include a new \tool fuzzer stage, and the
ability to fixup inputs after resizing.
Finally, since \work{NestFuzz} does not have a \work{FuzzBench} configuration,
we use the configuration provided in the project README.
\begin{table}
  \centering
  \begin{tabular}{llll}
    \toprule
    Type & Ref & Name & Version \\
    \midrule
    Binary & \cite{afl++} & \work{AFL++} & \texttt{v4.21c} \\
    Binary & \cite{libafl} & \work{LibAFL} & \texttt{f343376} \\
    Binary & \cite{afl} & \work{AFL} & \texttt{v2.57b} \\
    \midrule
    Structured & \cite{nestfuzz} & \work{NestFuzz} & \texttt{d16eb69} \\
    Structured & \cite{weizz} & \work{WEIZZ} & \texttt{c9cbeef} \\
    \bottomrule \\
  \end{tabular}
  \caption{Fuzzers used in our evaluation.}
  \label{table:fuzzers}
\end{table}

\customsection{Hardware}
We ran the large-scale fuzzing experiment on Google Cloud C3 instances with Intel Sapphire Rapids processors. The case-studies ran on dedicated servers with two Intel(R) Xeon(R) Gold 6430 @ 3.40GHz and 1 TB of RAM.

\customsection{Coverage Measurement}
We evaluate fuzzer performance by running each resulting corpus through a build of each benchmark instrumented with LLVM coverage, computing the total edge coverage.

\subsection{RQ1: Fuzzing Performance}
\label{sec:rq1}

To understand \tool's performance compared to
state-of-the-art fuzzers, we ran a large-scale fuzzing experiment with 16
benchmarks\footnote{Applying \work{NestFuzz} to \texttt{bloaty} and
\texttt{openssl} requires non-trivial build modifications that we were unable to
make.} and 7 fuzzers/fuzzer configurations. We tested fuzz runs from both an
empty corpus (denoted E in the results tables; here, ability to learn structure
quickly is particularly important) and
from a corpus with a single high-quality seed (denoted S in the results tables; here, ability to find variants of
the seed quickly is important).

We ran each fuzzer/benchmark/corpus configuration 10 times for 48 hours. We report the arithmetic mean edge coverage after 48 hours in \autoref{table:cov_table}.

Additionally, we compute an average score for each fuzzer, following the
conventions of \work{FuzzBench}.
For each target, we compute a fuzzer's score as the
percentage of maximum coverage obtained (where maximum coverage is the
\emph{highest} coverage obtained by any fuzzer).
For example, if a fuzzer
achieved the most coverage on a benchmark, it has a score of 100 for that
benchmark. If the fuzzer achieves only 70\% of the maximum coverage for that
benchmark, it receives a score of 70. The mean scores are reported in
\autoref{table:cov_table} (last row).

\customsection{Results}
\autoref{table:cov_table} shows results. 
For both both the empty and seeded corpus configurations, the \tool variants
were the most effective fuzzers, achieving the highest coverage in 10/16 of the
benchmarks in both cases. \work{AFL++(FS)} achieved the highest average
score (97.3) in the empty corpus by a margin of more than 7 points, followed by
\work{LibAFL(FS)} (89.6). In the seeded corpus setting, \work{LibAFL(FS)} achieved the
highest average score (96.7), followed by \work{LibAFL} (95.1) and then \work{AFL++(FS)} (94.1). This
ordering mirrors the baseline fuzzers themselves, where \work{AFL++} performs
(relatively) better from an empty corpus, while \work{LibAFL} performs better
from a seeded corpus.

The other three fuzzers (\work{AFL}, \work{WEIZZ}, and \work{NestFuzz}) were
generally not competitive, finding the highest coverage on only three benchmarks
across both corpus configurations; in fact, these three fuzzers underperform
the baseline fuzzers.
This is likely because our evaluation uses state-of-the-art
\work{FuzzBench} configurations.

\begin{table*}
  \centering
  \small
  \begin{tabularx}{\linewidth}{X|r@{\hspace{6pt}}rr@{\hspace{6pt}}r|r@{\hspace{6pt}}rr@{\hspace{6pt}}r|r@{\hspace{6pt}}rr@{\hspace{6pt}}rr@{\hspace{6pt}}r}
\toprule
& \multicolumn{2}{c}{\textbf{\work{AFL++(FS)}}} & \multicolumn{2}{c|}{\textbf{\work{LibAFL(FS)}}} & \multicolumn{2}{c}{\textbf{\work{AFL++}}} & \multicolumn{2}{c|}{\textbf{\work{LibAFL}}} & \multicolumn{2}{c}{\textbf{\work{AFL}}} & \multicolumn{2}{c}{\textbf{\work{WEIZZ}}} & \multicolumn{2}{c}{\textbf{\work{NestFuzz}}} \\
\cmidrule(lr){2-3} \cmidrule(lr){4-5} \cmidrule(lr){6-7} \cmidrule(lr){8-9} \cmidrule(lr){10-11} \cmidrule(lr){12-13} \cmidrule(lr){14-15}
Benchmark & \multicolumn{1}{c}{E} & \multicolumn{1}{c}{S} & \multicolumn{1}{c}{E} & \multicolumn{1}{c}{S} & \multicolumn{1}{c}{E} & \multicolumn{1}{c}{S} & \multicolumn{1}{c}{E} & \multicolumn{1}{c}{S} & \multicolumn{1}{c}{E} & \multicolumn{1}{c}{S} & \multicolumn{1}{c}{E} & \multicolumn{1}{c}{S} & \multicolumn{1}{c}{E} & \multicolumn{1}{c}{S} \\
\midrule
bloaty & 2095 & 2468 & \textbf{2330} & \textbf{4551} & 1858 & 1904 & 2005 & 3825 & 643 & 3157 & 732 & 959 & $\dagger$ & $\dagger$ \\
freetype2 & 8242 & 10142 & 8456 & 10491 & \textbf{9441} & \textbf{10715} & 8985 & 10317 & 3796 & 7572 & 4301 & 5093 & 3719 & 7284 \\
harfbuzz & 6835 & \textbf{7124} & 6661 & 6810 & \textbf{7034} & 7046 & 6688 & 6992 & 4020 & 5309 & 3243 & 3878 & 3950 & 5156 \\
lcms & \textbf{1940} & \textbf{2093} & 1852 & 2070 & 1184 & 1810 & 1753 & 2084 & 1250 & 1203 & 1601 & 1571 & 36 & 551 \\
libjpeg & \textbf{1739} & 2344 & 587 & 2288 & 950 & \textbf{2367} & 510 & 2318 & 656 & 2299 & 451 & 1948 & 478 & 2316 \\
libpcap & \textbf{3041} & \textbf{3010} & 2623 & 2846 & 2835 & 2678 & 2468 & 2689 & 36 & 2497 & 2094 & 2095 & 35 & 2068 \\
libpng & \textbf{1943} & \textbf{1991} & 1798 & 1965 & 1860 & 1963 & 1804 & 1962 & 1549 & 1936 & 1270 & 1690 & 7 & 1191 \\
ms-tpm-20-ref & 2845 & 3180 & \textbf{2889} & \textbf{3253} & 2209 & 2696 & 2685 & 3047 & 2409 & 2868 & 2072 & 2281 & 2029 & 2223 \\
openh264 & 8498 & 8491 & 8386 & 8305 & 8469 & 8485 & 8414 & 8457 & 8502 & \textbf{8521} & 6976 & 7178 & \textbf{8509} & 8496 \\
openssl & \textbf{4670} & \textbf{5178} & 4506 & 4906 & 4007 & 4879 & 4644 & 4690 & 4593 & 4773 & 3654 & 4299 & $\dagger$ & $\dagger$ \\
openthread & 2424 & 2505 & 2614 & \textbf{3000} & 2495 & 2521 & \textbf{2654} & 2962 & 2287 & 2380 & 2444 & 2655 & 2239 & 2373 \\
qpdf & 1181 & 1999 & 1157 & 1851 & \textbf{1203} & \textbf{2228} & 1169 & 1886 & 1165 & 1574 & 947 & 969 & 440 & 1045 \\
vorbis & \textbf{954} & 1253 & 509 & 1239 & 945 & 1264 & 397 & 1253 & 205 & 1257 & 206 & 1246 & 205 & \textbf{1267} \\
woff2 & \textbf{939} & \textbf{1060} & 934 & 1043 & 769 & 1043 & 814 & 1043 & 7 & 1003 & 713 & 1011 & 7 & 990 \\
\midrule
jsoncpp \scriptsize{(text)} & \textbf{510} & \textbf{510} & 507 & 508 & 509 & \textbf{510} & 507 & 507 & 508 & 508 & 508 & 508 & 508 & 157 \\
libxml2 \scriptsize{(text)} & 12551 & 12392 & 13130 & 13265 & 12838 & \textbf{14069} & \textbf{13251} & 13200 & 12366 & 10934 & 7442 & 7460 & 7107 & 7289 \\
\midrule
\textit{Average Score} & \textbf{97.3} & 94.1 & 89.6 & \textbf{96.7} & 88.6 & 92.1 & 86.6 & 95.1 & 61.7 & 84.4 & 63.9 & 71.9 & 42.4 & 68.8 \\
\bottomrule
  \end{tabularx}
  \vspace{5pt}
  \caption{Arithmetic mean edge coverage after 48\,h (10 runs) for each fuzzer--benchmark pair. Highest average coverage for each benchmark is in bold (for both empty and seeded corpus). E: empty corpus, S: seeded corpus. $\dagger$: target failed to build.}
  \label{table:cov_table}
\end{table*}

\subsection{RQ2: Applicability}
\label{sec:rq2}

\newcolumntype{F}[0]{>{\centering\arraybackslash}m{0.58in}}

\begin{table*}
  \centering
  \small
  \begin{tabularx}{\linewidth}{X|Frl|Frl|Frl|Frl}
    \toprule
    & \multicolumn{6}{c|}{\textbf{\work{AFL++(FS)} vs. \work{AFL++}}} & \multicolumn{6}{c}{\textbf{\work{LibAFL(FS) vs. LibAFL}}} \\
    \cmidrule(lr){2-7} \cmidrule(lr){8-13}
    Benchmark & Empty & \multicolumn{1}{c}{$\Delta$\%} & \multicolumn{1}{c|}{$p$} & Seeded & \multicolumn{1}{c}{$\Delta$\%} & \multicolumn{1}{c|}{$p$} & Empty & \multicolumn{1}{c}{$\Delta$\%} & \multicolumn{1}{c|}{$p$} & Seeded & \multicolumn{1}{c}{$\Delta$\%} & \multicolumn{1}{c}{$p$} \\[0.5ex]
    \hline
    libjpeg & \cellcolor{green!15}\includegraphics{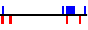} & \cellcolor{green!15}\textcolor{ForestGreen}{+83.0} & \cellcolor{green!15}\textcolor{ForestGreen}{\scriptsize{*}} & \includegraphics{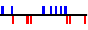} & -0.9 & {\scriptsize{}} & \includegraphics{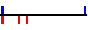} & +15.1 & {\scriptsize{}} & \includegraphics{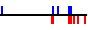} & -1.3 & {\scriptsize{}} \\
    lcms & \cellcolor{green!15}\includegraphics{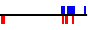} & \cellcolor{green!15}\textcolor{ForestGreen}{+63.9} & \cellcolor{green!15}\textcolor{ForestGreen}{\scriptsize{**}} & \cellcolor{green!15}\includegraphics{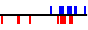} & \cellcolor{green!15}\textcolor{ForestGreen}{+15.6} & \cellcolor{green!15}\textcolor{ForestGreen}{\scriptsize{*}} & \includegraphics{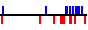} & +5.7 & {\scriptsize{}} & \includegraphics{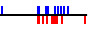} & -0.7 & {\scriptsize{}} \\
    bloaty & \cellcolor{green!15}\includegraphics{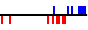} & \cellcolor{green!15}\textcolor{ForestGreen}{+12.7} & \cellcolor{green!15}\textcolor{ForestGreen}{\scriptsize{**}} & \cellcolor{green!15}\includegraphics{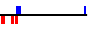} & \cellcolor{green!15}\textcolor{ForestGreen}{+29.6} & \cellcolor{green!15}\textcolor{ForestGreen}{\scriptsize{*}} & \includegraphics{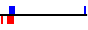} & +16.2 & {\scriptsize{}} & \includegraphics{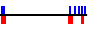} & +19.0 & {\scriptsize{}} \\
    ms-tpm-20-ref & \cellcolor{green!15}\includegraphics{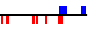} & \cellcolor{green!15}\textcolor{ForestGreen}{+28.8} & \cellcolor{green!15}\textcolor{ForestGreen}{\scriptsize{***}} & \cellcolor{green!15}\includegraphics{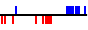} & \cellcolor{green!15}\textcolor{ForestGreen}{+18.0} & \cellcolor{green!15}\textcolor{ForestGreen}{\scriptsize{***}} & \cellcolor{green!15}\includegraphics{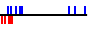} & \cellcolor{green!15}\textcolor{ForestGreen}{+7.6} & \cellcolor{green!15}\textcolor{ForestGreen}{\scriptsize{**}} & \cellcolor{green!15}\includegraphics{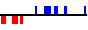} & \cellcolor{green!15}\textcolor{ForestGreen}{+6.7} & \cellcolor{green!15}\textcolor{ForestGreen}{\scriptsize{***}} \\
    woff2 & \cellcolor{green!15}\includegraphics{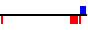} & \cellcolor{green!15}\textcolor{ForestGreen}{+22.1} & \cellcolor{green!15}\textcolor{ForestGreen}{\scriptsize{***}} & \cellcolor{green!15}\includegraphics{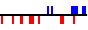} & \cellcolor{green!15}\textcolor{ForestGreen}{+1.6} & \cellcolor{green!15}\textcolor{ForestGreen}{\scriptsize{**}} & \cellcolor{green!15}\includegraphics{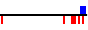} & \cellcolor{green!15}\textcolor{ForestGreen}{+14.8} & \cellcolor{green!15}\textcolor{ForestGreen}{\scriptsize{***}} & \includegraphics{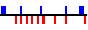} & +0.1 & {\scriptsize{}} \\
    libpcap & \cellcolor{green!15}\includegraphics{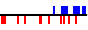} & \cellcolor{green!15}\textcolor{ForestGreen}{+7.3} & \cellcolor{green!15}\textcolor{ForestGreen}{\scriptsize{**}} & \cellcolor{green!15}\includegraphics{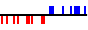} & \cellcolor{green!15}\textcolor{ForestGreen}{+12.4} & \cellcolor{green!15}\textcolor{ForestGreen}{\scriptsize{***}} & \includegraphics{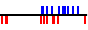} & +6.3 & {\scriptsize{}} & \cellcolor{green!15}\includegraphics{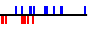} & \cellcolor{green!15}\textcolor{ForestGreen}{+5.8} & \cellcolor{green!15}\textcolor{ForestGreen}{\scriptsize{**}} \\
    openssl & \cellcolor{green!15}\includegraphics{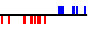} & \cellcolor{green!15}\textcolor{ForestGreen}{+16.5} & \cellcolor{green!15}\textcolor{ForestGreen}{\scriptsize{***}} & \cellcolor{green!15}\includegraphics{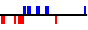} & \cellcolor{green!15}\textcolor{ForestGreen}{+6.1} & \cellcolor{green!15}\textcolor{ForestGreen}{\scriptsize{**}} & \cellcolor{red!15}\includegraphics{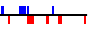} & \cellcolor{red!15}\textcolor{BrickRed}{-3.0} & \cellcolor{red!15}\textcolor{BrickRed}{\scriptsize{**}} & \cellcolor{green!15}\includegraphics{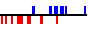} & \cellcolor{green!15}\textcolor{ForestGreen}{+4.6} & \cellcolor{green!15}\textcolor{ForestGreen}{\scriptsize{**}} \\
    vorbis & \includegraphics{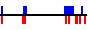} & +1.0 & {\scriptsize{}} & \cellcolor{red!15}\includegraphics{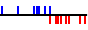} & \cellcolor{red!15}\textcolor{BrickRed}{-0.9} & \cellcolor{red!15}\textcolor{BrickRed}{\scriptsize{***}} & \includegraphics{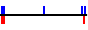} & +28.2 & {\scriptsize{}} & \cellcolor{red!15}\includegraphics{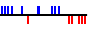} & \cellcolor{red!15}\textcolor{BrickRed}{-1.1} & \cellcolor{red!15}\textcolor{BrickRed}{\scriptsize{***}} \\
    libpng & \includegraphics{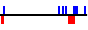} & +4.4 & {\scriptsize{}} & \cellcolor{green!15}\includegraphics{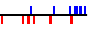} & \cellcolor{green!15}\textcolor{ForestGreen}{+1.4} & \cellcolor{green!15}\textcolor{ForestGreen}{\scriptsize{**}} & \includegraphics{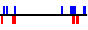} & -0.3 & {\scriptsize{}} & \includegraphics{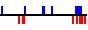} & +0.1 & {\scriptsize{}} \\
    jsoncpp & \includegraphics{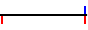} & +0.0 & {\scriptsize{}} & \includegraphics{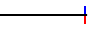} & +0.0 & {\scriptsize{}} & \includegraphics{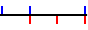} & +0.0 & {\scriptsize{}} & \cellcolor{green!15}\includegraphics{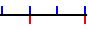} & \cellcolor{green!15}\textcolor{ForestGreen}{+0.2} & \cellcolor{green!15}\textcolor{ForestGreen}{\scriptsize{*}} \\
    openh264 & \cellcolor{green!15}\includegraphics{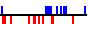} & \cellcolor{green!15}\textcolor{ForestGreen}{+0.3} & \cellcolor{green!15}\textcolor{ForestGreen}{\scriptsize{*}} & \includegraphics{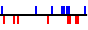} & +0.1 & {\scriptsize{}} & \includegraphics{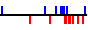} & -0.3 & {\scriptsize{}} & \cellcolor{red!15}\includegraphics{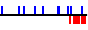} & \cellcolor{red!15}\textcolor{BrickRed}{-1.8} & \cellcolor{red!15}\textcolor{BrickRed}{\scriptsize{**}} \\
    openthread & \includegraphics{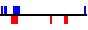} & -2.8 & {\scriptsize{}} & \includegraphics{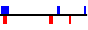} & -0.6 & {\scriptsize{}} & \includegraphics{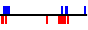} & -1.5 & {\scriptsize{}} & \includegraphics{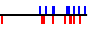} & +1.3 & {\scriptsize{}} \\
    harfbuzz & \cellcolor{red!15}\includegraphics{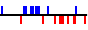} & \cellcolor{red!15}\textcolor{BrickRed}{-2.8} & \cellcolor{red!15}\textcolor{BrickRed}{\scriptsize{*}} & \includegraphics{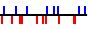} & +1.1 & {\scriptsize{}} & \includegraphics{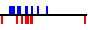} & -0.4 & {\scriptsize{}} & \cellcolor{red!15}\includegraphics{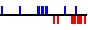} & \cellcolor{red!15}\textcolor{BrickRed}{-2.6} & \cellcolor{red!15}\textcolor{BrickRed}{\scriptsize{**}} \\
    libxml2 & \includegraphics{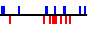} & -2.2 & {\scriptsize{}} & \cellcolor{red!15}\includegraphics{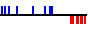} & \cellcolor{red!15}\textcolor{BrickRed}{-11.9} & \cellcolor{red!15}\textcolor{BrickRed}{\scriptsize{***}} & \includegraphics{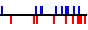} & -0.9 & {\scriptsize{}} & \includegraphics{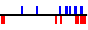} & +0.5 & {\scriptsize{}} \\
    qpdf & \includegraphics{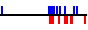} & -1.9 & {\scriptsize{}} & \cellcolor{red!15}\includegraphics{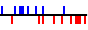} & \cellcolor{red!15}\textcolor{BrickRed}{-10.3} & \cellcolor{red!15}\textcolor{BrickRed}{\scriptsize{**}} & \includegraphics{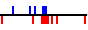} & -1.0 & {\scriptsize{}} & \includegraphics{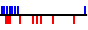} & -1.8 & {\scriptsize{}} \\
    freetype2 & \cellcolor{red!15}\includegraphics{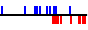} & \cellcolor{red!15}\textcolor{BrickRed}{-12.7} & \cellcolor{red!15}\textcolor{BrickRed}{\scriptsize{**}} & \cellcolor{red!15}\includegraphics{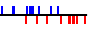} & \cellcolor{red!15}\textcolor{BrickRed}{-5.3} & \cellcolor{red!15}\textcolor{BrickRed}{\scriptsize{**}} & \includegraphics{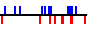} & -5.9 & {\scriptsize{}} & \includegraphics{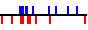} & +1.7 & {\scriptsize{}} \\[-0.5ex]
\bottomrule
    \end{tabularx}
    \vspace{5pt}
    \caption[\tool coverage compared to baseline fuzzers. Each graphic shows the final coverage of 10 \tool runs compared to 10 baseline fuzzer runs after 48 hours. \tool runs are represented by blue bars on top of the centerline, baseline fuzzer runs are red bars beneath the centerline. The left side of the scale represents the lowest coverage obtained by any run, the right side represents the most, scaled linearly. The average coverage change with \tool is shown in the $\Delta\%$ column. $p$ contains the Mann-Whitney U test p-value for statistical significance: *: $p < 0.05$, **: $p < 0.01$, ***: $p <0.001$]{\tool coverage compared to baseline fuzzers. Each graphic shows the final coverage of 10 \tool runs compared to 10 baseline fuzzer runs after 48 hours. \tool runs are represented by blue bars on top of the centerline, baseline fuzzer runs are red bars beneath the centerline. The left side of the scale represents the lowest coverage obtained by any run, the right side represents the most, scaled linearly. The average coverage change with \tool is shown in the $\Delta\%$ column. $p$ contains the Mann-Whitney U test p-value for statistical significance: *: $p < 0.05$, **: $p < 0.01$, ***: $p <0.001$ \\
    \includegraphics{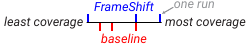}
    }
    \label{table:sota_delta}
\end{table*}

To understand the specific contribution of \tool over the baseline fuzzers \work{AFL++} and \work{LibAFL},
we visualize the final coverage values for each of the 10 fuzzer runs per benchmark/fuzzer in \autoref{table:sota_delta}. Each graphic shows the \tool-enabled variant runs (blue lines above the centerline) along with the baseline fuzzer runs (red lines below the centerline). The lines are plotted on a linear axis where the left-most side represents the run with the least coverage, and the right-most side represents the run with the most coverage. For each configuration, we report the average change in coverage due to enabling \tool ($\Delta\%$). Following best practices for fuzzer evaluations~\cite{klees2018evaluating, schloegel2024sok}, we use the Mann-Whitney U-test to compute the statistical significance of differences in fuzzer performance. These results are indicated in the $p$ column, as: * ($p < 0.05$), ** ($p < 0.01$), or *** ($p < 0.001$). Statistically significant results (using $p < 0.05$) are shaded green (\tool found more coverage) or red (\tool found less coverage) depending on the change in coverage.

\customsection{Results}
The direct baseline comparison is broken down more granularly in \autoref{table:sota_delta}. Of the two variants, we find that the \work{AFL++} variant gets more utility from the \tool integration, with a statistically significant increase in coverage in 15/32 configurations, and a decrease in only 6 configurations. For \work{LibAFL}, the effect is more muted: a statistically significant increase in 6 configurations and a decrease in 4. In both cases, the magnitudes of the coverage increase are generally larger than the decrease.

Aggregating the data, there are 7 benchmarks where \tool obtains a statistically significant increase in coverage of at least 3\% (\texttt{libjpeg}, \texttt{lcms}, \texttt{bloaty}, \texttt{ms-tpm-20-ref}, \texttt{woff2}, \texttt{libpcap}, \texttt{openssl}), 6 benchmarks where it is roughly neutral (\texttt{vorbis}, \texttt{libpng}, \texttt{jsoncpp}, \texttt{openh264}, \texttt{openthread}, \texttt{harfbuzz}), and 3 benchmarks where it has a negative effect (\texttt{libxml2}, \texttt{qpdf}, and \texttt{freetype2}).

All of the 7 positive benchmarks contain serialized size and/or offset fields that \tool is able to identify, thus enabling the baseline fuzzer to find differently-sized variants more quickly, contributing to finding coverage more quickly. Generally, there are two cases. 1. \tool enables rapid discovery of core coverage: all of the \tool runs end up near the highest found coverage, while baseline fuzzer results are more distributed (for example, \texttt{woff2}/\work{LibAFL}/Empty). Or 2. \tool enables breakout coverage discovery: one or more of the \tool runs is able to find significantly more coverage due to unlocking a certain codepath (e.g. \texttt{bloaty}/\work{AFL++}/Seeded).

Of the neutral benchmarks, several include serialized size/offset fields, yet obtain minimal changes in coverage. In \texttt{libpng} and \texttt{openthread} for example, it appears as though \tool is useful in the first few hours of fuzzing, but the baseline variants catch up after 48 hours.

We find an interesting case of \textit{frameshift-resistant} file formats, where \tool is not able to identify any fields, and thus fails to be productive. In both \texttt{vorbis} and \texttt{openh264}, the expected file format contains \textit{sync markers}, explicitly intended to prevent \textit{frameshift} issues when the file is  streamed across an unreliable medium. Thus, the parsers can recover when data is improperly resized (for example due to packet loss), and continue parsing mostly unharmed. As a result, \tool does not directly observe \textit{frameshifts} in the first part of the \textit{double-mutant} experiment, thus does not identify relation fields.

The case most adversarial to \tool is when the target generates an extremely large corpus and/or large files. In \texttt{harfbuzz}, \texttt{libxml2}, \texttt{qpdf}, and \texttt{freetype2}, the generated corpora are an order of magnitude larger than other benchmarks (tens of thousands of files), thus \tool spends more time analyzing the corpus and less time fuzzing. Even though \tool can identify relations in \texttt{harfbuzz} and \texttt{freetype2}, it is burdened by the analysis overhead. For text formats, the inability to find relation fields does not directly confer a negative performance (as demonstrated by \texttt{jsoncpp}), however coupled with the analysis overhead of a large corpus, it may reduce the amount of time available to fuzz (as in \texttt{libxml2}).

\customsection{Takeaways}
Given these results, a fuzzing practitioner could likely benefit from enabling \tool in most binary formats, especially when there is no available seed corpus. Incompatible formats, where \tool does not find relation fields will likely not directly confer a negative effect, unless the size of the corpus also grows too quickly. However, for industry-length fuzz campaigns (on the order of weeks), the effect of analysis overhead would be reduced as the corpus begins to saturate. For shorter fuzz campaigns, an interesting future direction could be to detect quick corpus growth and selectively apply \tool, balancing the fuzz time vs. analysis time.

\subsection{RQ3: Structure Recovery}
\label{sec:rq3}

In this section we qualitatively demonstrate case study examples that
demonstrate \tool's ability to identify relation fields in real-world formats. 

\subsubsection{PNG}
\label{sec:rq2-png}

\begin{figure}
  \resizebox{\linewidth}{!}{
    \includegraphics{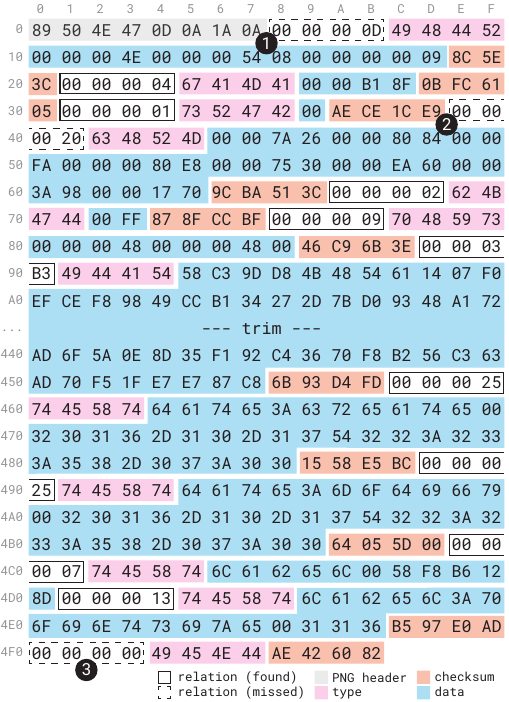}
  }
  \caption{Relation fields identified by \tool in a PNG file using \texttt{libpng}.}
  \label{fig:png}
\end{figure}

In \autoref{fig:png} we show the fields \tool finds in a PNG file using
\texttt{libpng}. \tool took only 313 milliseconds to analyze this file and
invoked the target 4705 times during the search.

The PNG file consists of an 8-byte PNG header followed by chunks of data. Each
chunk has a 4-byte size, 4-byte header (pink), N-byte data (blue), and 4-byte
checksum (gray). \tool correctly identifies 9 relation fields in the input (solid black outline).
Each of the relation fields it
identifies has the correct target span (the subsequent data portion of the
file).

\tool did not identify three potentially-expected relation fields, namely the
sizes for the initial \lt{IHDR} \inlinecircle{1}, \lt{cHRM} \inlinecircle{2}, and final \lt{IEND} \inlinecircle{3} chunks.
While these locations are identified as plausible candidates (and may be labeled
a size field by other approaches, such as those relying on manual grammars), \tool
eliminates them during the insertion point discovery phase
because it is unable to generate a \textit{restorative mutant} after changing their size.
Analyzing the code, we find that these chunks are validated
to be a fixed size during processing. Any size other than 13
(\lt{0xd}) for the \lt{IHDR} chunk will cause \texttt{libpng} to abort
early. In practice,
\lt{IHDR} is not resizable, and therefore \tool does not learn a relation for it. The same
holds for the \lt{cHRM} and \lt{IEND} chunks. 
That is, although superficially similar to the other size fields in this input, these fields do not actually describe parts of the input which can be resized.

\subsubsection{ASN.1}
\label{sec:rq2-asn1}

\begin{figure}
  \resizebox{\linewidth}{!}{
    \includegraphics{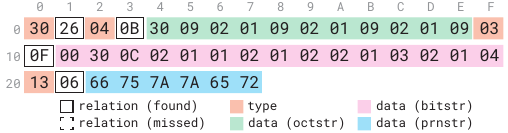}
  }
  \caption{Relation fields identified by \tool in a DER-encoded ASN.1 file file using \texttt{openssl}.}
  \label{fig:asn1}
\end{figure}

\autoref{fig:asn1} shows the relation fields \tool finds in a DER-encoded ASN.1 file when run against the \texttt{asn1} tool in \texttt{openssl}. This input took 59 milliseconds to analyze and required 63 invocations of the target.

The file represents a nested \lt{SEQUENCE} object
containing three entries: an octet string with
values \lt{[9,9,9]}, a bitstring with values
\lt{[1,2,3,4]}, and a printable string with the value
\lt{\"fuzzer\"}.
Objects in the file are encoded in a TLV (type-length-value) format. Each object
contains a single byte type, followed by a multi-byte length, and then an
N-sized value. 

\tool identifies the outer sequence length (at \lt{0x01}) and all three of the inner object lengths (at \lt{0x03}, \lt{0x10}, and \lt{0x21}).
In this format, length fields are variable-size: values smaller than 127 are encoded in short form, as a single byte. However, larger values use a prefix byte \lt{0x80 + x} where $x$ indicates how many additional bytes are in the encoding for the length field.

\tool does not naturally support this variable-size length field construction, yet is still able to approximate the relation field by identifying only the low-order bytes. Specifically, for
single-byte values, the field \emph{appears} to be a single-byte value until
the size boundary is crossed.
Similarly, in larger cases, \tool
interprets the encoded field \lt{[82 03 12]} (value: 0x123) as a
2-byte big endian integer---not 3!---covering only \lt{[03 12]}. As long as mutations don't
cause this size to require one more or fewer byte to represent, it serves as an
accurate representation of the real field, thus most fuzzer mutations result in an accurate adjustment. In practice, even though \tool is imprecise, it still enables a statistically significant increase in coverage on \texttt{openssl} by more than 6\%.

\subsubsection{ELF}

\begin{figure}
  \resizebox{\linewidth}{!}{
    \includegraphics{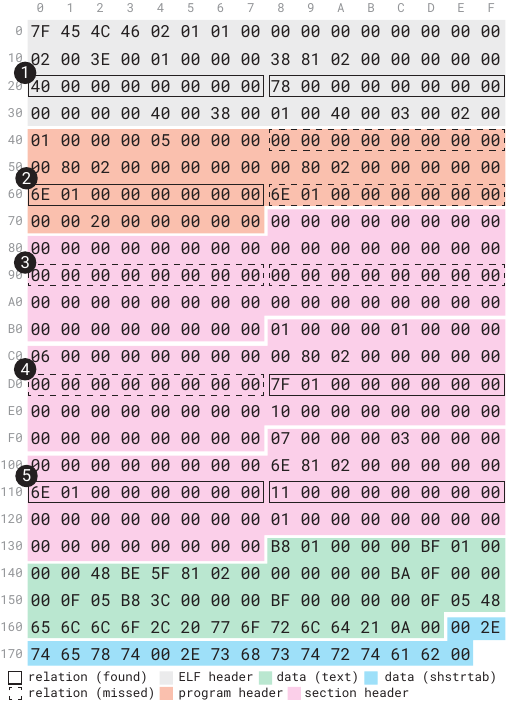}
  }
  \caption{Relation fields identified by \tool in an ELF file using \texttt{bloaty}.}
  \label{fig:elf}
\end{figure}

In \autoref{fig:elf}, we visualize the relation fields \tool identified in an ELF file with the \texttt{bloaty} target, which parses several types of object files. This file took 458 milliseconds to analyze and \tool invoked the target 1571 times during the search.

The file consists of an ELF header (grey, \lt{0-0x40}), a program table header with one entry (orange, \lt{0x40-0x78}), a section table header with three entries (pink, \lt{0x78-0x138}), and some data that is mapped into the \lt{.text} (green, \lt{0x138-0x16E}) and the \lt{.shstrtab} (blue, \lt{0x16E-0x17F}) sections.

In the ELF header, \tool correctly identifies the offset fields \inlinecircle{1} for both the program header table (at \lt{0x20}) and the section header table (at \lt{0x28}). Note that \tool must actually discover the section header table \textit{first} because insertions before the program header require shifting both offsets.

The program header table contains a single entry, which in turn contains fields \inlinecircle{2} representing the size on disk (at \lt{0x60}) and in memory (at \lt{0x68}) of a segment to map, along with the file offset (at \lt{0x48}). \tool identifies the file size but not the offset (because the ELF header is required to start at the beginning) or the size in memory, because \texttt{bloaty} only parses the file and does not attempt to execute it.

The first section \inlinecircle{3} is a necessary \lt{null} section, and thus while there are size/offset fields according to the spec, they are not used and \tool does not identify them. The next section is the \lt{.text} section \inlinecircle{4}. This section describes both the offset (at \lt{0xD0}) and length (at \lt{0xD8}) of code data to map. Here the section actually maps the entire file starting from the beginning through the green region, which contains machine code. \tool does not identify the offset field because moving the ELF header caused corruption, but it did identify the size field with the correct parameters.

The third section \inlinecircle{5} describes the \texttt{.shstrtab} section which points to a list of strings that are used to identify the section names. It also contains both an offset (at \lt{0x110}) and size (at \lt{0x118}). Here \tool identifies both the offset and the indirect size field.

\customsection{Bonus: Automatic Rebase}
While analyzing this ELF example, we were pleasantly surprised to find that \tool identified fields accurately enough able to perform a non-trivial rebase operation. In general, \emph{resizing} parts of an ELF file is difficult, as it requires changing considerable metadata,  and typically is only accomplished by dedicated tools, such as \texttt{lief}~\cite{LIEF}.

However, \tool is able to discover enough information (in less than half a second) to accurately fix up the metadata for mutations that splice the contents of the code region (green). Normally, splicing here would corrupt the binary considerably (as several fields are out of sync); with \tool, however, we performed a splice mutation to insert new machine code into this region, shrinking its size. \tool automatically updated one offset field (at \lt{0x110}) and two size fields (at \lt{0xD0} and \lt{0x60}), preserving the validity of the file. The resulting file correctly ran as an executable without any additional modifications!

While this is a small example, it serves to demonstrate that \tool is capable of automatically identifying the important relations in structures and performing potentially-complex size-changing mutations while preserving validity. Thus, in practice, \tool is quite effective on the \texttt{bloaty} target, achieving a statistically significant increase in coverage of more than 20\% over the baseline fuzzers.

\subsection{RQ4: Versatility}
\label{sec:rq4}

\tool makes few assumptions about the underlying framework or
type of coverage feedback (unlike systems like \work{NestFuzz} which depend on specific toolchains). Additionally, its modular integration into existing fuzzers allows it to naturally extend baseline capabilities, such as \work{LibAFL}'s support for targets besides C/C++.  We therefore demonstrate \tool's versatility by applying it to two case study target programs in other languages: Rust (Section~\ref{sec:rust}) and Python (Section~\ref{sec:python}). For these experiments we utilize our \work{LibAFL} prototype of \tool.

We chose to apply \tool to two programs which parse similar file formats to some of our C/C++ benchmark programs (\texttt{PNG} in Rust, and \texttt{ASN.1} in Python) to compare the similarity between the types of structures \tool can identify and illustrate how even programs which parse the same format can have semantic differences.

\subsubsection{\tool in Rust: PNG}
\label{sec:rust}

Our first target was the Rust crate \texttt{image-png},\footnote{https://github.com/image-rs/image-png} a pure Rust library for image decoding and encoding. The project has several \texttt{cargo-fuzz} harnesses,
including a \lt{decode} harness that decodes arbitrary bytes as a PNG file.
We compile the target with sanitizer coverage (as we do for C/C++), since the Rust
build system uses LLVM internally.

We ran the \autoref{fig:png} test PNG
to determine which structures \tool could learn in this Rust program.
It took 1.7 seconds to run and required invoking the target 9933 times.
Note that a structure-inference tool performing any amount of static
analysis would likely incur overhead from moving to Rust,
since the compiler produces boilerplate that can obfuscate
program logic.
Since \tool is fully dynamic, any additional overhead comes from runtime
performance, which is often negligible if the target is compiled with optimizations.

\tool found several different fields when fuzzing the
Rust (\texttt{image-png}) and C (\texttt{libpng}) image libraries.  Interestingly, these differences are associated with true behavioral differences
in the target programs.
In the Rust program, \tool ignores the \lt{IHDR} chunk size (as in \texttt{libpng}), since this is parsed as a fixed-size chunk. It finds the next two length fields (at \lt{0x21} and \lt{0x31}), but then it \textit{also} identifies the \lt{cHRM} size field \inlinecircle{2}, unlike with \texttt{libpng}.
This is because the semantics of PNG actually differ
between \texttt{libpng} and \texttt{image-png}:
while \texttt{libpng} immediately validates the size of \lt{cHRM} to be 32 bytes, \texttt{image-png} has no such check. Thus, in \texttt{image-png} it \emph{is} resizable, and \tool correctly identifies this.

\tool identifies the next two size fields (at \lt{0x6A} and \lt{0x78}), but not the size field of the big \lt{IDAT} chunk at \lt{0x8D}, nor any of the following chunk sizes. Although initially confused, we discovered that \texttt{image-png}'s \lt{decode} harness does \emph{not} actually parse the whole \texttt{PNG} file. The fuzz harness instead keeps iterating over chunks until it has found all the actual image data (stored in \lt{IDAT} chunks). In this case, the file contains just a single \lt{IDAT} chunk. Thus, \texttt{image-png} scans through the file until reading this chunk and then exits early (ignoring the remaining comment chunks).

From a \texttt{PNG} grammar perspective, these comment chunks have size fields. Yet in the context of this harness, they are ignored and can be freely mutated without inducing a loss in coverage (in fact they do not contribute coverage to begin with); thus \tool (rightly) does not consider them to be relation fields.

\subsubsection{\tool in Python: ASN.1}
\label{sec:python}

We also evaluated \tool on a Python case study, which uses a different compiler toolchain and alternative coverage feedback.
Such an adaptation would be fundamentally quite difficult for a static analysis-based approach without significant effort. 
\tool supports this out-of-the-box, using \work{LibAFL's} existing support for \work{Atheris}~\cite{google_atheris_2023},
a coverage-guided Python fuzzer.
Python represents a higher abstraction level than C/C++, e.g.,  list concatenation in a C++ program may hit many basic blocks (i.e. invoking stdlib functions), while the same operation corresponds a single Python opcode in Python. This changes the  nature of coverage instrumentation.

Our target program was \texttt{pyasn1},\footnote{https://github.com/etingof/pyasn1}
a Python-based framework for encoding and decoding ASN.1 files. We used the same ASN.1 file from \autoref{fig:asn1} as a seed.
Running the analysis took only 25 milliseconds and 125 invocations of the target program.
\tool found the same fields as it did on the \texttt{openssl} benchmark.
In this case, however, it incorrectly identified the byte at position \lt{0x20} as a size field.
This is because, when \tool corrupted this field and performed a nearby insertion,
it ended up coincidentally achieving enough of the same coverage features (through some other mechanism);
therefore, \tool categorized the field as a relation.
Increasing the $t_{loss}$ threshold was sufficient to remove this false positive,
which suggests that \tool's parameters may benefit from tuning when applying it
to frameworks with different types of coverage feedback.

\section{Discussion}

Our evaluation demonstrates that relation-field structure inference is both possible to perform using only standard greybox coverage and \textit{practical} enough that it can be incorporated into state-of-the-art fuzzers for a net improvement in coverage. \tool is capable of identifying relation structures in many real world formats and its relaxed assumptions about the type of coverage and lack of additional instrumentation allow it to be integrated easily into other applications, such as fuzzing Rust and Python programs.

\tool serves an example of a potentially more general class of techniques which use coverage feedback in conjunction with heuristics to search the space of possible input structures and augment fuzzers. This approach need not be limited just to size and offset fields. For example, a similar double-mutant experiment may be able to identify  compressed/encoded regions of the input (e.g. \texttt{zlib}, \texttt{base64}, etc.) or repeatable parts of the input (i.e. chunks). A key challenge, as we have seen first-hand in prototyping these kinds of ideas, is keeping the time required for analysis sufficiently tractable to produce wins in the resulting structure-aware fuzzing.  

Another interesting future direction, however, could augment this type of coverage-guided inference with a more powerful heuristic, such as a large language model, which could be used to propose high-quality candidate structures. It is possible that such a technique could discover much more complex (and more useful) structures, that could be validated with dynamic experimentation; the additional complexity might effectively counterbalance additional required analysis time. 

Additionally, it may be possible to automatically tune certain parameters of \tool based on the target program. For example, limiting analysis time if the corpus grows too quickly, or tuning the $t_{loss}$ threshold for different types of coverage feedback.

\section{Related Work}

\customsection{Structure inference for binary fuzzing}
Most similar to \tool are tools which performs automatic structure-inference to aid fuzzers. \work{NestFuzz}~\cite{nestfuzz} performs a dynamic taint analysis (DTA) to learn the input processing logic for a program. \work{TIFF}~\cite{tiff} uses DTA to infer the types of various input fields. Most recently, \work{AIFORE}~\cite{aifore} fused byte-level taint analysis with machine learning for byte-level clustering.

\work{WEIZZ}~\cite{weizz} and \work{ProFuzzer}~\cite{profuzzer} take a more greybox approach (like us) using just instrumentation to guide inference. While \work{ProFuzzer} uses existing coverage feedback, \work{WEIZZ} requires not just the final coverage, but the \emph{order} of hitting certain bits. While \work{ProFuzzer} performs a similar type of coverage-destroying analysis (like the first step in our double-mutant experiment), it does not attempt to restore coverage through a second mutation, labeling the byte a ``size field'' without understanding the relation to other fields. Neither is capable of learning both size/offset fields \textit{and} understanding what parts of the input they describe.

\customsection{Structure inference for reverse-engineering}
A parallel body of work recovers \emph{internal} structures to aid decompilation or static analysis~\cite{tie,tupni,discoverer,lin2008deriving,howard,lin2010automatic}.
While conceptually similar, these approaches target the structures used internally to the program not the serialized structure of the input. Thus, their results are not immediately useful for fuzzer mutations.

\customsection{Specification-based fuzzing}
An alternative approach to fuzzing binary formats provides the fuzzer with a specification beforehand. \work{AFLSmart}~\cite{aflsmart} can perform smart chunk-based mutations when provided with the virtual structure of a file format. These mutations are similar in essence to the types of mutations \tool can enable (in that they set size fields accurately), relying on the existence of a manual specification. \work{FormatFuzzer}~\cite{formatfuzzer} repurposes structure format files used by a file structure explorer utility, converting them into C++ programs which can \textit{generate} and \textit{mutate} instances of the format. Similarly, the \work{ISLa}~\cite{isla} project aims to create an input specification language that can be sampled using a constraint solver. These approaches are interesting and useful when such a format is available. However, they can be onerous to provide. Beyond the initial complexity of the task, programs may parse multiple formats at once (i.e. \texttt{bloaty}), or implement the semantics of a given format \textit{differently} than another program that nominally does the same thing:
while \tool can learn program-specific formats (as it did with \texttt{PNG} in \texttt{libpng} vs. \texttt{image-png}), these specification-driven fuzzers cannot.

\customsection{Grammar-based fuzzing}
Structure-aware fuzzing has been more heavily utilized in the context of fuzzing text-based formats such as scripting language interpreters~\cite{jsfunfuzz, holler2012fuzzing, ifuzzer, codealchemist, nautilus, zhou2023towards, godefroid2008grammar}. In contrast to structured binary formats, these text-based formats are usually representable with (or can be approximated by) a context-free grammar (CFG). These formats typically do not contain serialized size or offset fields, and thus do not suffer from the same \textit{frameshift} problem as binary formats. 

Tools like \work{GLADE}~\cite{glade}, \work{Pygmalion}~\cite{pygmalion}, \work{Skyfire}~\cite{skyfire}, and \work{Autogram}~\cite{autogram} try to automatically learn such input grammars.
Given existing grammars, fuzzers like \work{Nautilus}~\cite{nautilus} and \work{Gramatron}~\cite{gramatron} can perform \textit{coverage-guided} semantic-preserving mutations.
Perhaps the most related work in this other domain is \work{Grimoire}~\cite{grimoire} which upon receiving a new input tries to understand how to \textit{generalize} the input by observing how different mutations change the coverage bitmap.

\section{Conclusion}

Destructive frameshift mutations remain a central obstacle to effective coverage-guided fuzzing of binary formats, inhibiting fuzzers from exploring the space of valid inputs. \tool mitigates this problem by learning size- and offset-relations directly from standard coverage feedback and by preserving those relations during mutation. The approach requires no manual specification and integrates transparently with \work{AFL++} and \work{LibAFL}.

In a 12 CPU-year evaluation, across 16 real-world benchmarks, we show that \tool raises edge coverage by an average of 6 percent--and sometimes more than 50 percent--outperforming five other state-of-the-art fuzzers. Further, the approach is language-agnostic and we successfully applied it (with no modifications!) to both Rust and Python.

\tool thus offers a practical solution to combat the \textit{frameshift} problem--complementing the existing performance of modern fuzzers while allowing them to learn to resize inputs without breaking them.

\bibliographystyle{plain}
\bibliography{references}

\begin{thebibliography}{10}

\bibitem{nautilus}
Cornelius Aschermann, Tommaso Frassetto, Thorsten Holz, Patrick Jauernig, Ahmad-Reza Sadeghi, and Daniel Teuchert.
\newblock Nautilus: Fishing for deep bugs with grammars.
\newblock In {\em NDSS}, 2019.

\bibitem{redqueen}
Cornelius Aschermann, Sergej Schumilo, Tim Blazytko, Robert Gawlik, and Thorsten Holz.
\newblock Redqueen: Fuzzing with input-to-state correspondence.
\newblock In {\em NDSS}, volume~19, pages 1--15, 2019.

\bibitem{discord2024:eqv_tlv}
Cornelius (eqv /~.eqv) Aschermann.
\newblock {“something that I've seen commonly that made fuzzing VERY difficult is TLV kind of formats, where the header has an overall size field , and each chunk has a size field, and we have sizeof(header)+header.size + sum(chunks.size) == file\_size and sum(chunks.size) == header.size” — Discord message in \emph{\#general} (Awesome Fuzzing server)}, September 2024.
\newblock Message ID 1287160766469767199; posted 2024-09-21 17:17; accessed 2025-06-05.

\bibitem{glade}
Osbert Bastani, Rahul Sharma, Alex Aiken, and Percy Liang.
\newblock Synthesizing program input grammars.
\newblock {\em ACM SIGPLAN Notices}, 52(6):95--110, 2017.

\bibitem{grimoire}
Tim Blazytko, Matt Bishop, Cornelius Aschermann, Justin Cappos, Moritz Schl{\"o}gel, Nadia Korshun, Ali Abbasi, Marco Schweighauser, Sebastian Schinzel, Sergej Schumilo, et~al.
\newblock $\{$GRIMOIRE$\}$: Synthesizing structure while fuzzing.
\newblock In {\em 28th USENIX Security Symposium (USENIX Security 19)}, pages 1985--2002, 2019.

\bibitem{discoverer}
Weidong Cui, Jayanthkumar Kannan, and Helen~J Wang.
\newblock Discoverer: Automatic protocol reverse engineering from network traces.
\newblock In {\em USENIX Security Symposium}, pages 1--14. Boston, MA, USA, 2007.

\bibitem{tupni}
Weidong Cui, Marcus Peinado, Karl Chen, Helen~J Wang, and Luis Irun-Briz.
\newblock Tupni: Automatic reverse engineering of input formats.
\newblock In {\em Proceedings of the 15th ACM conference on Computer and communications security}, pages 391--402, 2008.

\bibitem{nestfuzz}
Peng Deng, Zhemin Yang, Lei Zhang, Guangliang Yang, Wenzheng Hong, Yuan Zhang, and Min Yang.
\newblock Nestfuzz: Enhancing fuzzing with comprehensive understanding of input processing logic.
\newblock In {\em Proceedings of the 2023 ACM SIGSAC Conference on Computer and Communications Security}, pages 1272--1286, 2023.

\bibitem{formatfuzzer}
Rafael Dutra, Rahul Gopinath, and Andreas Zeller.
\newblock Formatfuzzer: Effective fuzzing of binary file formats.
\newblock {\em ACM Transactions on Software Engineering and Methodology}, 33(2):1--29, 2023.

\bibitem{weizz}
Andrea Fioraldi, Daniele~Cono D'Elia, and Emilio Coppa.
\newblock Weizz: Automatic grey-box fuzzing for structured binary formats.
\newblock In {\em Proceedings of the 29th ACM SIGSOFT international symposium on software testing and analysis}, pages 1--13, 2020.

\bibitem{afl++}
Andrea Fioraldi, Dominik Maier, Heiko Ei{\ss}feldt, and Marc Heuse.
\newblock $\{$AFL++$\}$: Combining incremental steps of fuzzing research.
\newblock In {\em 14th USENIX Workshop on Offensive Technologies (WOOT 20)}, 2020.

\bibitem{libafl}
Andrea Fioraldi, Dominik Maier, Dongjia Zhang, and Davide Balzarotti.
\newblock {LibAFL: A Framework to Build Modular and Reusable Fuzzers}.
\newblock In {\em Proceedings of the 29th ACM conference on Computer and communications security (CCS)}, CCS '22. ACM, November 2022.

\bibitem{godefroid2008grammar}
Patrice Godefroid, Adam Kiezun, and Michael~Y Levin.
\newblock Grammar-based whitebox fuzzing.
\newblock In {\em Proceedings of the 29th ACM SIGPLAN conference on programming language design and implementation}, pages 206--215, 2008.

\bibitem{google_atheris_2023}
Google.
\newblock {Atheris}: A coverage-guided, native python fuzzer.
\newblock \url{https://github.com/google/atheris}, 2023.
\newblock Version 2.3.0, commit cbf4ad9, accessed 2025-06-06.

\bibitem{pygmalion}
Rahul Gopinath, Bj{\"o}rn Mathis, Mathias H{\"o}schele, Alexander Kampmann, and Andreas Zeller.
\newblock Sample-free learning of input grammars for comprehensive software fuzzing.
\newblock {\em arXiv preprint arXiv:1810.08289}, 2018.

\bibitem{codealchemist}
HyungSeok Han, DongHyeon Oh, and Sang~Kil Cha.
\newblock Codealchemist: Semantics-aware code generation to find vulnerabilities in javascript engines.
\newblock In {\em NDSS}, 2019.

\bibitem{holler2012fuzzing}
Christian Holler, Kim Herzig, and Andreas Zeller.
\newblock Fuzzing with code fragments.
\newblock In {\em 21st $\{$USENIX$\}$ Security Symposium ($\{$USENIX$\}$ Security 12)}, pages 445--458, 2012.

\bibitem{autogram}
Matthias Hoschele and Andreas Zeller.
\newblock Mining input grammars with autogram.
\newblock In {\em 2017 IEEE/ACM 39th International Conference on Software Engineering Companion (ICSE-C)}, pages 31--34. IEEE, 2017.

\bibitem{tiff}
Vivek Jain, Sanjay Rawat, Cristiano Giuffrida, and Herbert Bos.
\newblock Tiff: using input type inference to improve fuzzing.
\newblock In {\em Proceedings of the 34th annual computer security applications conference}, pages 505--517, 2018.

\bibitem{klees2018evaluating}
George Klees, Andrew Ruef, Benji Cooper, Shiyi Wei, and Michael Hicks.
\newblock Evaluating fuzz testing.
\newblock In {\em Proceedings of the 2018 ACM SIGSAC conference on computer and communications security}, pages 2123--2138, 2018.

\bibitem{tie}
JongHyup Lee, Thanassis Avgerinos, and David Brumley.
\newblock Tie: Principled reverse engineering of types in binary programs.
\newblock 2011.

\bibitem{lin2008deriving}
Zhiqiang Lin and Xiangyu Zhang.
\newblock Deriving input syntactic structure from execution.
\newblock In {\em Proceedings of the 16th ACM SIGSOFT International Symposium on Foundations of software engineering}, pages 83--93, 2008.

\bibitem{lin2010automatic}
Zhiqiang Lin, Xiangyu Zhang, and Dongyan Xu.
\newblock Automatic reverse engineering of data structures from binary execution.
\newblock In {\em Proceedings of the 11th Annual Information Security Symposium}, pages 1--1, 2010.

\bibitem{liu2023sbft}
Dongge Liu, Jonathan Metzman, Marcel B{\"o}hme, Oliver Chang, and Abhishek Arya.
\newblock Sbft tool competition 2023-fuzzing track.
\newblock In {\em 2023 IEEE/ACM International Workshop on Search-Based and Fuzz Testing (SBFT)}, pages 51--54. IEEE, 2023.

\bibitem{fuzzbench}
Jonathan Metzman, L{\'a}szl{\'o} Szekeres, Laurent Simon, Read Sprabery, and Abhishek Arya.
\newblock Fuzzbench: an open fuzzer benchmarking platform and service.
\newblock In {\em Proceedings of the 29th ACM joint meeting on European software engineering conference and symposium on the foundations of software engineering}, pages 1393--1403, 2021.

\bibitem{miller1990empirical}
Barton~P Miller, Lars Fredriksen, and Bryan So.
\newblock An empirical study of the reliability of unix utilities.
\newblock {\em Communications of the ACM}, 33(12):32--44, 1990.

\bibitem{aflsmart}
Van-Thuan Pham, Marcel B{\"o}hme, Andrew~E Santosa, Alexandru~R{\u{a}}zvan C{\u{a}}ciulescu, and Abhik Roychoudhury.
\newblock Smart greybox fuzzing.
\newblock {\em IEEE Transactions on Software Engineering}, 47(9):1980--1997, 2019.

\bibitem{jsfunfuzz}
Jesse Ruderman.
\newblock Introducing jsfunfuzz.
\newblock {\em URL http://www. squarefree. com/2007/08/02/introducing-jsfunfuzz}, 20:25--29, 2007.

\bibitem{schloegel2024sok}
Moritz Schloegel, Nils Bars, Nico Schiller, Lukas Bernhard, Tobias Scharnowski, Addison Crump, Arash Ale-Ebrahim, Nicolai Bissantz, Marius Muench, and Thorsten Holz.
\newblock Sok: Prudent evaluation practices for fuzzing.
\newblock In {\em 2024 IEEE Symposium on Security and Privacy (SP)}, pages 1974--1993. IEEE, 2024.

\bibitem{ossfuzz}
Kostya Serebryany.
\newblock $\{$OSS-Fuzz$\}$-google's continuous fuzzing service for open source software.
\newblock 2017.

\bibitem{aifore}
Ji~Shi, Zhun Wang, Zhiyao Feng, Yang Lan, Shisong Qin, Wei You, Wei Zou, Mathias Payer, and Chao Zhang.
\newblock $\{$AIFORE$\}$: Smart fuzzing based on automatic input format reverse engineering.
\newblock In {\em 32nd USENIX Security Symposium (USENIX Security 23)}, pages 4967--4984, 2023.

\bibitem{howard}
Asia Slowinska, Traian Stancescu, and Herbert Bos.
\newblock Howard: A dynamic excavator for reverse engineering data structures.
\newblock In {\em NDSS}, 2011.

\bibitem{gramatron}
Prashast Srivastava and Mathias Payer.
\newblock Gramatron: Effective grammar-aware fuzzing.
\newblock In {\em Proceedings of the 30th acm sigsoft international symposium on software testing and analysis}, pages 244--256, 2021.

\bibitem{isla}
Dominic Steinh{\"o}fel and Andreas Zeller.
\newblock Input invariants.
\newblock In {\em Proceedings of the 30th ACM joint european software engineering conference and symposium on the foundations of software engineering}, pages 583--594, 2022.

\bibitem{LIEF}
Romain Thomas.
\newblock Lief - library to instrument executable formats.
\newblock https://lief.quarkslab.com/, apr 2017.

\bibitem{ifuzzer}
Spandan Veggalam, Sanjay Rawat, Istvan Haller, and Herbert Bos.
\newblock Ifuzzer: An evolutionary interpreter fuzzer using genetic programming.
\newblock In {\em European Symposium on Research in Computer Security}, pages 581--601. Springer, 2016.

\bibitem{skyfire}
Junjie Wang, Bihuan Chen, Lei Wei, and Yang Liu.
\newblock Skyfire: Data-driven seed generation for fuzzing.
\newblock In {\em 2017 IEEE Symposium on Security and Privacy (SP)}, pages 579--594. IEEE, 2017.

\bibitem{profuzzer}
Wei You, Xueqiang Wang, Shiqing Ma, Jianjun Huang, Xiangyu Zhang, XiaoFeng Wang, and Bin Liang.
\newblock Profuzzer: On-the-fly input type probing for better zero-day vulnerability discovery.
\newblock In {\em 2019 IEEE symposium on security and privacy (SP)}, pages 769--786. IEEE, 2019.

\bibitem{afl}
Michal Zalewski.
\newblock American fuzzy lop.
\newblock \url{https://lcamtuf.coredump.cx/afl/}.
\newblock Accessed: September 10, 2024.

\bibitem{zhou2023towards}
Chijin Zhou, Quan Zhang, Lihua Guo, Mingzhe Wang, Yu~Jiang, Qing Liao, Zhiyong Wu, Shanshan Li, and Bin Gu.
\newblock Towards better semantics exploration for browser fuzzing.
\newblock {\em Proceedings of the ACM on Programming Languages}, 7(OOPSLA2):604--631, 2023.

\end{thebibliography}

\end{document}